\documentclass[aps,pre,twocolumn,nofootinbib,floatfix]{revtex4-1}
\usepackage[utf8]{inputenc}
\usepackage[T1]{fontenc}
\usepackage{lmodern,graphicx}
\usepackage{times}
\usepackage{mathptmx}
\usepackage{epstopdf}
\usepackage{xcolor}
\usepackage{graphicx}
\usepackage[colorlinks,linkcolor=red,urlcolor=blue,citecolor=red]{hyperref}
\usepackage[normalem]{ulem}
\usepackage{xspace}
\usepackage{float}
\usepackage{bm}
\hypersetup{
    colorlinks=true,
    linkcolor=blue,
    filecolor=magenta,      
    urlcolor=cyan,
    citecolor=cyan
}
\begin{document}
\title{Kuramoto model with additional nearest-neighbor interactions:
Existence of a nonequilibrium tricritical point}
\author{Mrinal Sarkar}
\affiliation{Department of Physics, Indian Institute of Technology
Madras, Chennai 600036, India}
\email{mrinal@physics.iitm.ac.in}
\author{Shamik Gupta}
\affiliation{\mbox{Department of Physics, Ramakrishna Mission Vivekananda Educational and
Research Institute, Belur Math, Howrah 711202, India}\\ \mbox{Regular Associate, Quantitative Life Sciences Section, ICTP -
The Abdus Salam International Centre for Theoretical Physics,}\\\mbox{Strada
Costiera 11, 34151 Trieste, Italy}}
\email{shamikg1@gmail.com}
\begin{abstract}
A paradigmatic framework to study the phenomenon of spontaneous
collective synchronization is provided by the Kuramoto model comprising
a large collection of phase oscillators of distributed frequencies
that are globally coupled through the sine of their phase differences. We study here a variation of the model by including
nearest-neighbor interactions on a one-dimensional lattice. While the
mean-field interaction resulting from the global coupling favors
global synchrony, the nearest-neighbor interaction may have cooperative
or competitive effects depending on the sign and the magnitude of the
nearest-neighbor coupling. For unimodal and symmetric frequency distributions, we demonstrate that as a result, the model
in the stationary state exhibits in contrast to the usual Kuramoto
model both continuous and first-order transitions between synchronized
and incoherent phases, with the transition lines meeting at a
tricritical point. Our results are based on numerical integration of the
dynamics as well as an approximate theory involving appropriate averaging of fluctuations in the stationary state.  
\end{abstract}
\maketitle
\section{Introduction}
\label{sec:intro}

Competing interactions are known to result in interesting stationary and
dynamical features in systems comprising many interacting degrees of
freedom. Here, we explore this theme within the ambit of a many-body
system involving {\it phase oscillators} of distributed natural
frequencies interacting via a mean-field
and a nearest-neighbor interaction on a one-dimensional periodic
lattice. In the absence of the nearest-neighbor interaction, the
dynamics is that of the Kuramoto model~\cite{kuramoto-book}, well known in the field of
nonlinear dynamics as a paradigmatic framework to study the phenomenon
of spontaneous synchronization abound in
nature~\cite{strogatzsync,pikovskybook}. The
model has been extensively employed over the years to explain the
emergence of collective synchrony in a diverse range of scenarios, from
 Josephson junction arrays~\cite{Wiesenfeld1998} and chemical
 oscillators~\cite{Taylor2009}, to
 power-grids~\cite{Taher2019}, rhythmic applause in concert
 halls~\cite{Nda2000}, and
 many more. 
 
 The dynamics of the Kuramoto model
is strictly non-Hamiltonian: it cannot be obtained as an
overdamped dynamics on a potential energy landscape, as is possible when
the natural frequencies are same for all the
oscillators. For unimodal and symmetric frequency distributions, the model in the limit of infinite system-size shows as a function
of the mean-field coupling a continuous phase transition between a
synchronized and an incoherent phase~\cite{kuramoto-book,strogatz2000}.
The former phase is characterized by a macroscopic number of oscillators
having different phases but nevertheless sharing a common frequency. In
the incoherent phase, however, there is no macroscopic cluster of
coherent oscillators. The Kuramoto model when considered with solely
nearest-neighbor interaction has been shown to not exhibit any
macroscopic phase locking and hence any synchronized phase on a
one-dimensional periodic lattice~\cite{Strogatz1988}.

In the aforementioned backdrop, we explore in this
work the issue of what happens when one includes both a mean-field and a
nearest-neighbor interaction in the Kuramoto setting. We show that as a
result, the system in the stationary state exhibits both synchronized
and incoherent phases; thus, the scenario of nonexistence of a synchronized phase
with solely nearest-neighbor interaction is significantly modified on
adding a mean-field interaction, in that the system now does exhibit a
synchronized phase. Moreover, a phase transition occurs
between the two phases as one tunes the relevant dynamical parameters,
with the transition being either continuous (with continuous variation
of the order parameter) or first-order (showing jumps in the behavior of
the order parameter at the transition point). The two transition
lines meet at a so-called tricritical point, defined as the termination of
a continuous transition and a first-order transition
point~\cite{Huang1987}. While existence of such
points has been demonstrated earlier for Hamiltonian systems relaxing to
equilibrium stationary states, see recent works, e.g., ~\cite{Barr2001,Antoniazzi2007}, our work is a demonstration of existence
of a tricritical point in a non-Hamiltonian dynamics relaxing to a
nonequilibrium stationary state, and is to the best of our knowledge a
hitherto unreported existence of such a point in the framework of the
Kuramoto model. An earlier demonstration of the existence of
a tricritical point in a nonequilibrium setting has been in the context
of stochastic dynamics of interacting many-particle
systems~\cite{soumen}, thus very
much different from the setup considered in this work. Our claims are supported by extensive
numerical integration results as well as an approximate theory valid in
the limit of large system size that considers an appropriate averaging
of fluctuations in the stationary state. 

The layout of the paper is as follows. In Section~\ref{sec:model}, we
define our model of study. In Section~\ref{sec:model-3}, we discuss a reparametrization
of the model convenient for further analysis, and list the main queries
addressed in this work. In
Section~\ref{sec:phase-diagram}, we present our results on the complete phase
diagram of the model, together with reporting on numerical
results that demonstrate the existence of both continuous and first-order
transitions in the stationary state of our model, and a discussion on how to obtain numerically the lines of continuous and
first-order transitions in the parameter space. In
Section~\ref{sec:tat}, we discuss an approximate theory to obtain the
order parameter variation in our model. The
paper ends with conclusions in Section~\ref{sec:conclusions}. In
Appendix \ref{app0}, we motivate our model from a perspective
different from that of interacting phase oscillators, namely, that of
classical rotors interacting via a mean-field and a nearest-neighbor
interaction which arises as a reduced model describing layered magnetic
structures. Appendix \ref{app1} provides a reminder of the scaling theory of continuous transitions in
equilibrium.

\section{Model and dynamics}
\label{sec:model}

\begin{figure}[]
\centering
\includegraphics[scale=0.45]{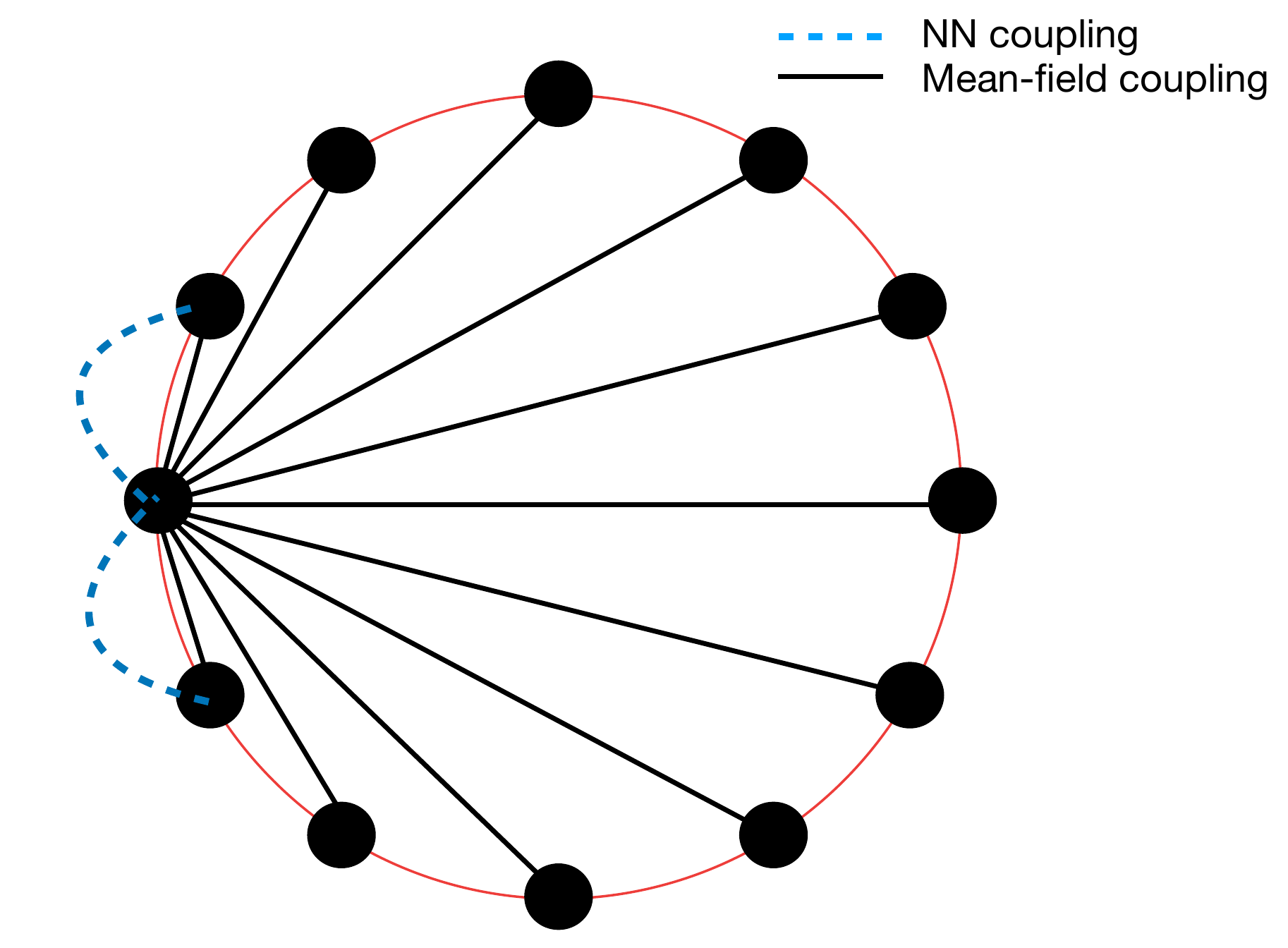}
\caption{(Color online) Schematic diagram showing the coupling scheme
for the model~(\ref{eq:eom1}) on a one-dimensional periodic lattice. The phase oscillators occupying the
lattice sites (black filled circles) have a mean-field and a nearest-neighbor coupling. For a
representative site, we have shown by black full lines (respectively, by
blue dashed lines) the mean-field (respectively, the nearest-neighbor
(NN)) coupling.}
\label{fig:coupling}
\end{figure}

We consider a one-dimensional periodic lattice of $L$ sites, with sites
labeled $i=1,2\ldots,L$. On each site resides a {\it phase oscillator}
interacting with oscillators on all other sites via a mean-field coupling
with strength $J$ and also with oscillators on its
nearest-neighbor sites with strength $K$. {
Figure~\ref{fig:coupling} shows the coupling scheme.} We take $J$ to be positive,
while $K$ can be of either sign.
Denoting by $\theta_i \in [0,2\pi);~\theta_{i+L}=\theta_i$ the angle~\cite{note-phase} of the oscillator on the $i$-th site,
the dynamics is defined by $L$ coupled nonlinear
differential equations of the form 
\begin{equation}
\frac{{\rm d} \theta_{i}}{{\rm d}t}=\omega_{i}+Jr\sin(\psi -
\theta_{i})+K\sum_{j \in nn_{i}}\sin(\theta_{j}- \theta_{i}).
\label{eq:eom1}
\end{equation}
Here, $\omega_i$ is the natural frequency of the $i$-th oscillator,
while the
second term on the right hand side (rhs) may be interpreted as a torque
(in suitable units) arising from a mean-field interaction and expressed in terms of the usual Kuramoto synchronization order
parameter~\cite{kuramoto-book,strogatz2000}
\begin{equation}
 r e^{{\rm i} \psi} \equiv \frac{1}{L}  \sum_{j=1}^{L} e^{{\rm i}
 \theta_{j}}.
\label{eq:order_para_definition}   
\end{equation}
On the other hand, the third term on the rhs of Eq.~(\ref{eq:eom1}) is the torque due to a nearest-neighbor interaction, with the sum over $j$ restricted to the nearest neighbors of $i$.
The $\omega_i$'s denote a set of quenched-disordered random variables
sampled independently from a common distribution $G(\omega)$ with finite
mean $\omega_0>0$ and width $\sigma>0$. The quantity $r;~0\le r \le 1$
in Eq.~(\ref{eq:order_para_definition}) is a
measure of the amount of synchrony present in the system at a given time
instant, while $\psi$ measures the average angle~\cite{strogatz2000}. As is usual in studies of the Kuramoto model, we
consider $G(\omega)$ to be unimodal, i.e., symmetric about $\omega_0$
and decreasing monotonically and continuously to zero with increasing
$|\omega-\omega_0|$. In view of rotational invariance of the
dynamics~(\ref{eq:eom1}), the effect of $\omega_0$ can be gotten rid of from
the dynamics by effecting the transformation $\theta_i \to
\theta_i+\omega_0 t~\forall~i$. On implementing such a transformation, one evidently has
$\omega_i$'s having zero mean in the resulting dynamics; we will from
now on consider such an implementation to have been made, and consider
instead of~(\ref{eq:eom1}) the dynamics
\begin{equation}
\frac{{\rm d} \theta_{i}}{{\rm d}t}=\sigma\omega_{i}+Jr\sin(\psi -
\theta_{i})+K\sum_{j \in nn_{i}}\sin(\theta_{j}- \theta_{i}).
\label{eq:eom1-1}
\end{equation}
Here, the $\omega_i$'s are now distributed according to a distribution
$g(\omega)$ that has zero mean and unit
variance. 

The dynamics~(\ref{eq:eom1-1}) is intrinsically non-Hamiltonian. This
may be understood as follows: although the torque due to the mean-field and the nearest-neighbor interaction may be obtained from a potential 
$V(\{\theta_i\})\equiv (J/2L)\sum_{i,j=1}^L
[1-\cos(\theta_i-\theta_j)]-K\sum_{i=1}^L
[\cos(\theta_{i+1}-\theta_i)+\cos(\theta_{i-1}-\theta_i)]$, a similar
procedure cannot be implemented for the frequency term. This is because
an ad hoc potential $\sim -\sum_{i=1}^L \sigma \omega_i \theta_i$
that would nevertheless allow to obtain the frequency term in the
dynamics~(\ref{eq:eom1-1}) would not be
periodic in the angle variables and thus cannot be regarded as a bona
fide potential of the system. As a result of the foregoing, the
dynamics~(\ref{eq:eom1-1}) cannot be interpreted as an overdamped
dynamics on a potential landscape, as is possible with $\omega_i=0
~\forall~ i$~\cite{Gupta2018}. In the latter case, the dynamics may be
written as
\begin{equation}
\frac{{\rm d}\theta_i}{{\rm d}t}=-\frac{\partial
V(\{\theta_i\})}{\partial \theta_i},
\label{eq:eom-overdamped}
\end{equation}
and then the long-time stationary solution corresponds to values of
$\theta_i$'s that minimize the potential
$V(\{\theta_j\})$~\cite{strogatz-book}. A consequence of the
non-Hamiltonian nature of the dynamics~(\ref{eq:eom1-1}) is that the
stationary state it relaxes to is not an equilibrium but rather a
nonequilibrium stationary
state~\cite{Gupta2018}. 

Setting $K$ to zero in Eq.~(\ref{eq:eom1-1}) recovers the usual
Kuramoto model that has only mean-field
interaction~\cite{kuramoto-book,acebron2005kuramoto,Gupta2014,Rodrigues2016,Gupta2018}, while setting $J$ to zero reduces the dynamics to the
version of the Kuramoto model with only nearest-neighbor
interaction~\cite{Strogatz1988}.
In the former case, it is known in the limit $L \to \infty$ that in the
stationary state, attained as $t\to \infty$, the model shows a continuous
phase transition from a low-$J$ incoherent phase (zero value of
the stationary $r$) to a high-$J$ synchronized
phase (a non-zero value for the stationary $r$) across the critical
point $J_c =2\sigma/(\pi g(0))$~\cite{strogatz2000,Gupta2018}. Study of the model with only
nearest-neighbor interaction has established that in the limit $L \to
\infty$, no angle locking and consequently, a non-zero value for
stationary $r$ is possible~\cite{Strogatz1988}.  

\section{Reparametrization of the dynamics and queries}
\label{sec:model-3}

For further analysis, we reduce the dynamics~(\ref{eq:eom1-1}) to a
dimensionless form. To this end, implementing for $J \ne 0$ the transformations $t \to
Jt,~\sigma \to \sigma/J,~K \to K/J$, one obtains the dimensionless form as  
\begin{equation}
\frac{{\rm d} \theta_{i}}{{\rm d}t} =  \sigma \omega_{i} + r \sin (\psi
- \theta_{i}) + K\sum_{j \in nn_{i}}  \sin(\theta_{j}- \theta_{i}).
\label{eq:eom2}
\end{equation}
{ The aforementioned transformations are tantamount to
considering the dynamics~(\ref{eq:eom1-1}) with $J=1$. We will show later
in this section that the relevant parameters to obtain phase transitions
in the dynamics~(\ref{eq:eom1-1}) are the ratios $\sigma/J$ and $K/J$, and hence, the
results on the order parameter variation when plotted, e.g., as a
function of $K/J$ and for a fixed $\sigma/J$, with different values of $J
\ne 0$, all coincide. The latter fact justifies the transformations
that have been invoked to rewrite the dynamics in the
form~(\ref{eq:eom2}).}  From now on, we will study the dynamics~(\ref{eq:eom2}) in
the parameter space $(\sigma,K)$. In obtaining numerical results
reported later in the paper, we employ as representative examples of the
frequency distribution a
Gaussian and a Lorentzian $g(\omega)$; $\sigma$ is
identified with the variance of the Gaussian distribution, and with the
half-width at half-maximum of the Lorentzian distribution.

 In the dimensionless dynamics~(\ref{eq:eom2}), the continuous
 transition of the usual Kuramoto model is observed as one tunes
 $\sigma$ across the critical value $\sigma_c=\pi g(0)/2$, with the
 system existing in the synchronized phase at low $\sigma$ and in the
 incoherent phase at high $\sigma$. In this backdrop, we ask: How does the inclusion
 of nearest-neighbor interaction modify the stationary-state phase diagram? Do new
 phases emerge? What is the order of transition between the different
 phases? We may anticipate new features in view of the fact that for
 $K<0$, the mean-field and nearest-neighbor
 interactions have competing tendencies: while the former favors
 global synchrony, the latter would like to make oscillator angles get
 out of phase on nearest-neighbor sites. For $K>0$, however, we expect
 both the mean-field and the local interaction to have cooperative
 effect in establishing global synchrony. In both the scenarios, an
 essential role will be played also by the parameter $\sigma$. In view of the
 foregoing, it is evidently pertinent to embark on a detailed
 analysis of the dynamics~(\ref{eq:eom2}), an issue we take up in this
 work. The results presented in the whole of
 Section~\ref{sec:phase-diagram} correspond to Gaussian $g(\omega)$,
 while the case of Lorentzian $g(\omega)$ is discussed in
 Section~\ref{sec:conclusions}. 

\section{Phase diagram of the model~(\ref{eq:eom2}) in ($\sigma-K$) plane}
\label{sec:phase-diagram}

\begin{figure}[]
\centering
\includegraphics[scale=0.45]{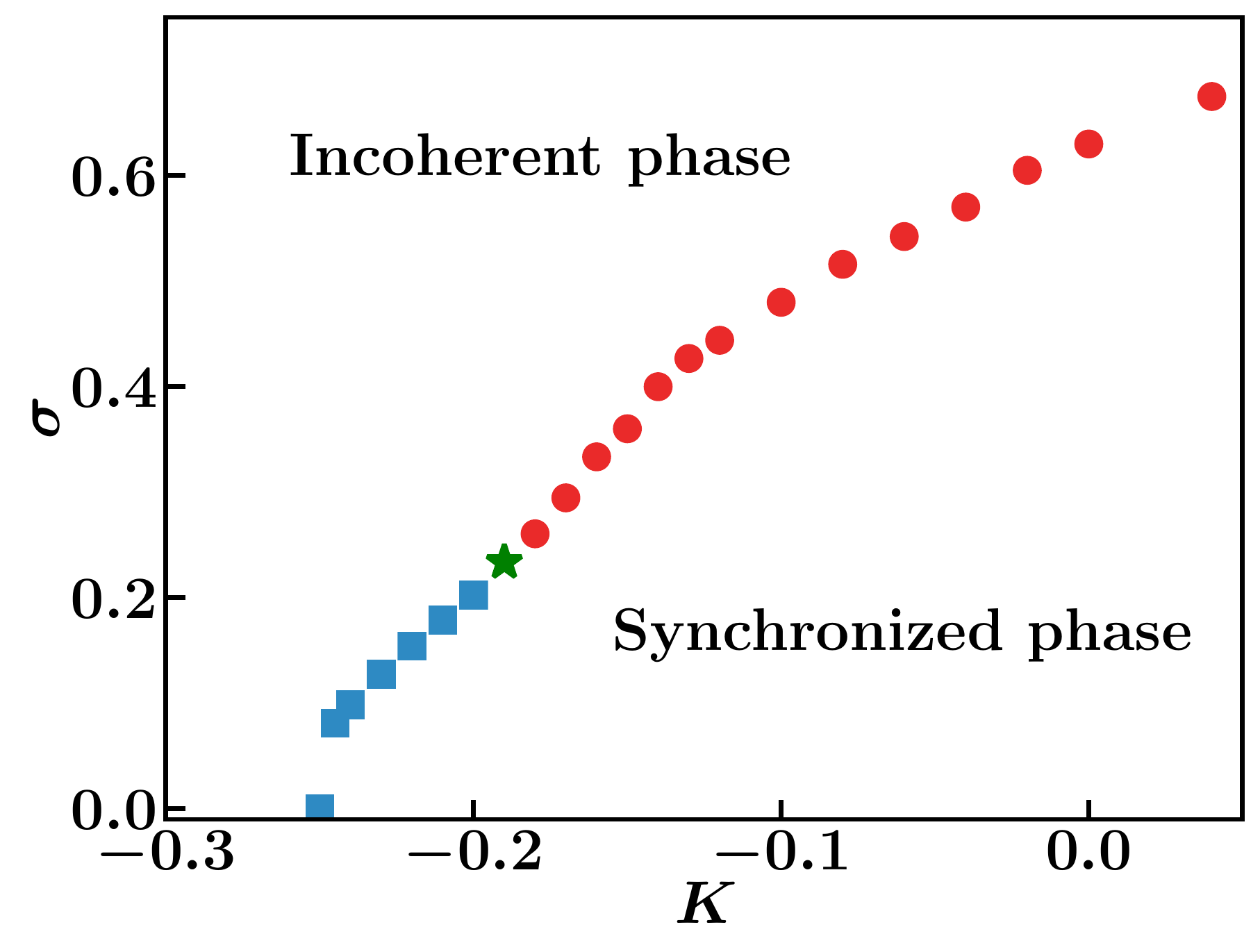}
\caption{(Color online) The complete phase diagram of the
model~(\ref{eq:eom2}) in the ($\sigma-K$) plane, showing synchronized and
incoherent phases separated by a line of transition that is either
first-order (blue squares) or continuous (red circles). The two lines meet at a tricritical point,
shown by a green star. The frequency distribution $g(\omega)$ is a
Gaussian with zero mean and unit variance. Exact results are obtained
for (i) $K=0$, yielding the critical point
$(\sigma_c=\sqrt{\pi}/(2\sqrt{2}),0)$, and (ii) $\sigma=0$, yielding the
critical point $(K_c=-0.25,\sigma=0)$.}
\label{fig:parameter_space}
\end{figure}

The stationary-state phase diagram of the model~(\ref{eq:eom2}) in the ($\sigma-K$) plane is shown in
Fig.~\ref{fig:parameter_space} for Gaussian $g(\omega)$, where the circles in red constitute
the line of continuous transition, while the line of first-order
transition is represented by squares in blue. The tricritical point is
located at ($\sigma_{\rm Tricritical} \approx 0.23,~K_{\rm
Tricritical} \approx -0.19$), and is denoted by a green star. We discuss
below how we obtain the phase diagram in Fig.~~\ref{fig:parameter_space} from numerical integration results
of the dynamics~(\ref{eq:eom2}) for large but finite $L$. For the system
sizes scanned, we did not observe any appreciable dependence of the
transition points on $L$. 

From the phase diagram, we see that for $K>0$, when both the mean-field
and the nearest-neighbor interaction favour global synchrony, one has a continuous phase transition
from a low-$\sigma$ synchronized phase to a high-$\sigma$ incoherent phase. For negative values of $K$, there is instead a competition between the two types of interaction. One has a
continuous transition as long as $K > K_{\rm Tricritical}$ and otherwise
a first-order transition. As stated earlier, for $K=0$, we recover the transition
point of the usual Kuramoto model.

For $\sigma=0$, we now discuss how one may obtain exact results for the
critical value $K_c$. In
this case, the dynamics~(\ref{eq:eom2}) takes the form of
Eq.~(\ref{eq:eom-overdamped}), with the potential in
dimensionless form given by 
\begin{eqnarray}
&&V(\{\theta_i\})= -r\sum_{i=1}^L\cos(\psi-\theta_i)\nonumber \\
&&-K\sum_{i=1}^L
[\cos(\theta_{i+1}-\theta_i)+\cos(\theta_{i-1}-\theta_i)].
\label{eq:potential}
\end{eqnarray}
As mentioned in Section~\ref{sec:model}, the stationary solution then
corresponds to values of $\theta_i$'s that minimize the potential $V$.
Consider first the incoherent phase, which has by definition a zero
value for stationary $r$, and the potential is minimized by having
angles of oscillators on nearest-neighbor sites differing by an amount
equal to $\pi$ (since $K$ is here negative, see
Fig.~\ref{fig:parameter_space}). The corresponding minimum value of the potential~(\ref{eq:potential}) is given by
\begin{equation}
V_{\rm inc}=2KL.
\end{equation}
On the other hand, the potential can also be minimized by having all the
angles equal to one another (which is the favored state for $\sigma=0$), yielding unity for the stationary $r$
(maximally synchronized phase) and
the potential having the corresponding value 
\begin{equation}
V_{\rm sync}=-L-2KL.
\end{equation}
It is then evident that equating $V_{\rm inc}$ with $V_{\rm sync}$
defines $K_c$ such that on either side of this critical value,
it is the incoherent or the synchronized phase that minimizes the potential
and is consequently observed in the stationary state. The equality
$2K_cL=-L-2K_cL$
yields the exact critical value $K_c=-0.25$.

The rest of this section is devoted to a detailed discussion of how one
may obtain the phase diagram in Fig.~\ref{fig:parameter_space} from an analysis of the dynamics~(\ref{eq:eom2}).

\subsection{Continuous versus first-order transitions}
\label{sec:continuous-vs-first-order}

\begin{figure}[]
\includegraphics[scale=0.4]{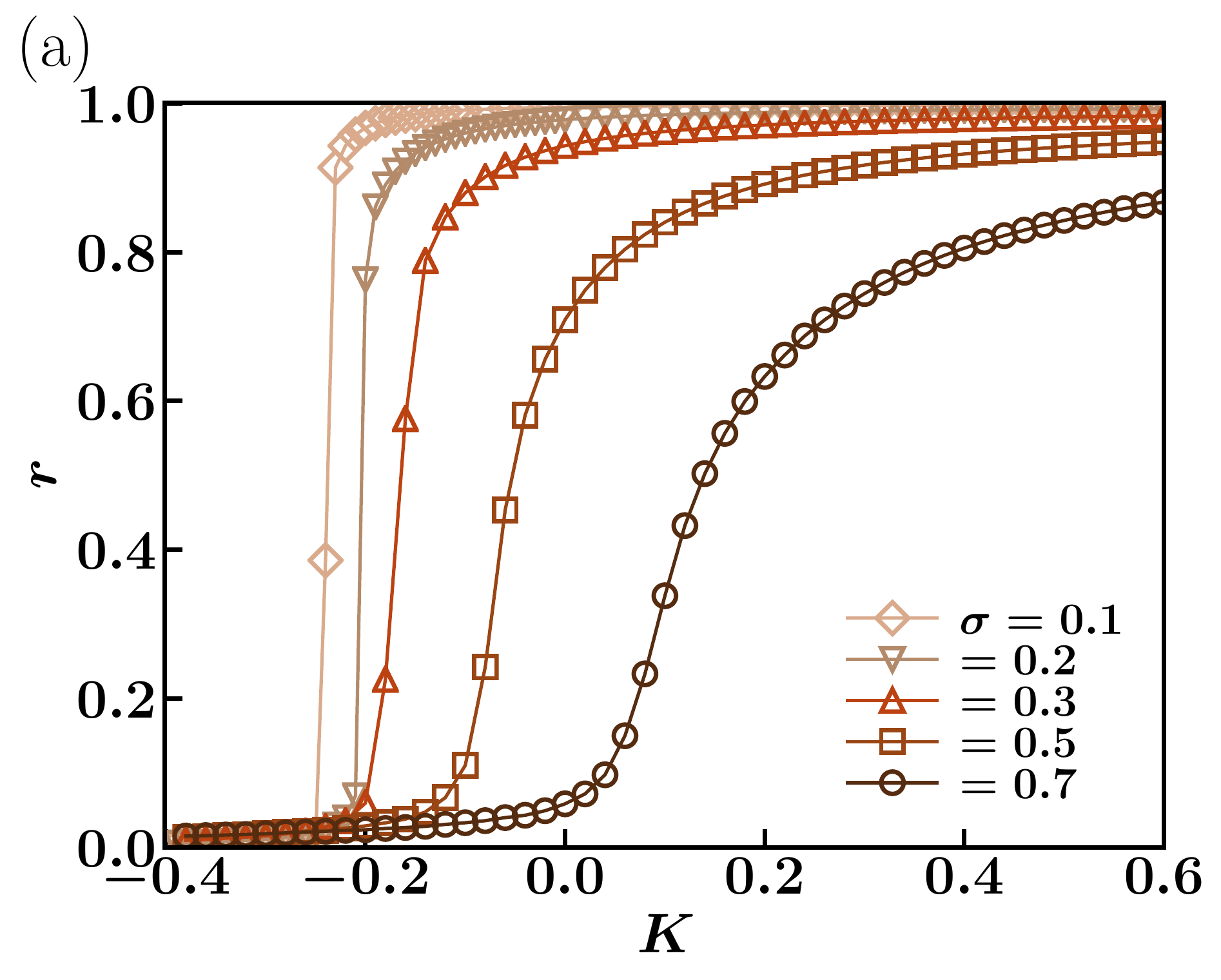}
\includegraphics[scale=0.4]{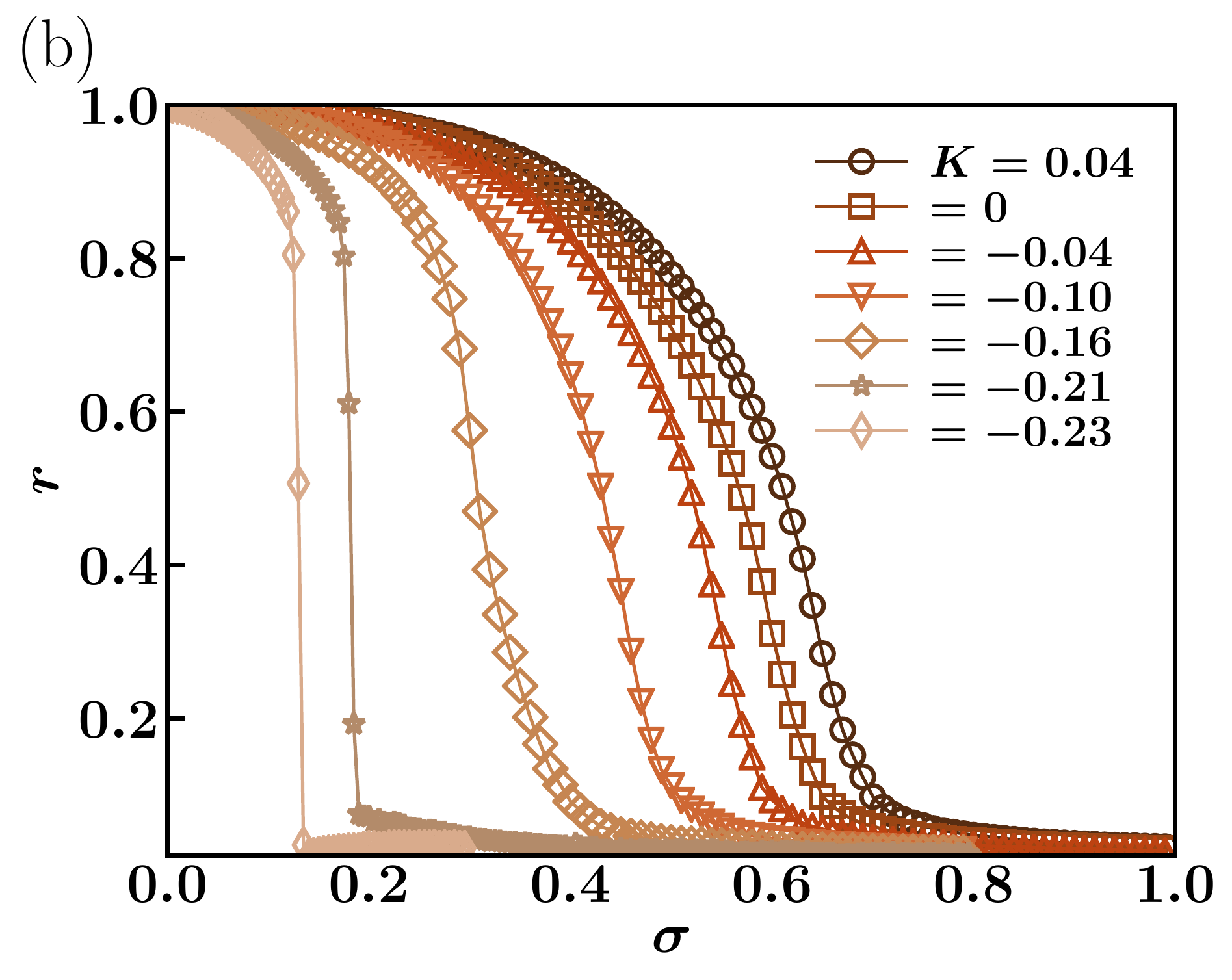}
\caption{(Color online) Variation of order parameter $r$ with $K$ for
several values of $\sigma$ (panel (a)) and that with $\sigma$ for
several values of $K$ (panel (b)) in the stationary state of the
dynamics~(\ref{eq:eom2}) on a lattice of size $L=3200$. The frequency distribution $g(\omega)$ is a
Gaussian with zero mean and unit variance.  The data have been averaged, first over dynamical
evolution in the stationary state for a given frequency realization
$\{\omega_i\}$, and then over different frequency realizations. 
Both the figures suggest the existence of both synchronized and incoherent phases and a
phase transition between them. The latter appears to be continuous
(continuous variation of $r$) for high values of $\sigma$, and to be
first-order-like (sharp jump of $r$) for low $\sigma$, as shown in panel (a). Similarly, the figure in panel (b) shows that
the transition from the synchronized to the incoherent phase appears
continuous for positive and low negative values of $K$ and first-order-like for large negative $K$. The data are obtained from numerical
integration of the dynamics~(\ref{eq:eom2}). In obtaining the results depicted
in the figure, we initiate, for every individual pair of values of
$\sigma$ and $K$, the dynamics~(\ref{eq:eom2}) in a state in which all the oscillators have the same angle; we then let
the system relax to stationarity, signalled by a time-independent value
of $r$, and record the latter value.}
\label{fig:rho_vs_sigma}
\end{figure}

In order to gain preliminary insights into possible dynamical behavior,
one may start off with performing numerical integration
of the dynamics~(\ref{eq:eom2}) by employing a fourth-order
Runge-Kutta algorithm with integration time step ${\rm d}t=0.01$ and for Gaussian $g(\omega)$.
Figure~\ref{fig:rho_vs_sigma}(a) shows for several values of $\sigma$ the
variation of the order parameter $r$ with $K$ in the stationary state on a lattice of
size $L=3200$~\cite{note-simulation}. In obtaining the results depicted
in the figure, we initiate, for every individual pair of values of $\sigma$
and $K$,  the dynamics~(\ref{eq:eom2}) in a state in which all the oscillators have the same angle; we then let
the system relax to stationarity, signalled by a time-independent value
of $r$, and record the latter value. Unless stated otherwise, the results for the order
parameter presented here and elsewhere in the paper have been obtained
by taking time average of the data in the stationary state for a given
frequency realization $\{\omega_i\}$ and considering a further average
over different frequency realizations. The figure suggests the existence of both synchronized and incoherent phases and a
phase transition between them. The latter appears to be continuous
(continuous variation) for high values of $\sigma$, and to be
first-order-like (sharp jump) for low $\sigma$.

Our phase diagram~\ref{fig:parameter_space} clearly shows that varying $K$ at a fixed $\sigma$
lets us reveal the nature of the phase transition in a way that is
completely equivalent to varying $\sigma$ at a fixed $K$. That this is
indeed the case is evident from the results presented in
Fig.~\ref{fig:rho_vs_sigma}(b) that shows for several values of $K$ the
variation of the order parameter $r$ with $\sigma$ in the stationary state on a lattice of
size $L=3200$~\cite{note-simulation}. Again, we see both synchronized
and incoherent phases, with a phase transition between them that appears to be continuous for positive
and low negative values of $K$, and to be first-order-like for large
negative $K$.

\begin{figure}[!ht]
\hspace{-0.52 cm}
\includegraphics[scale=0.22]{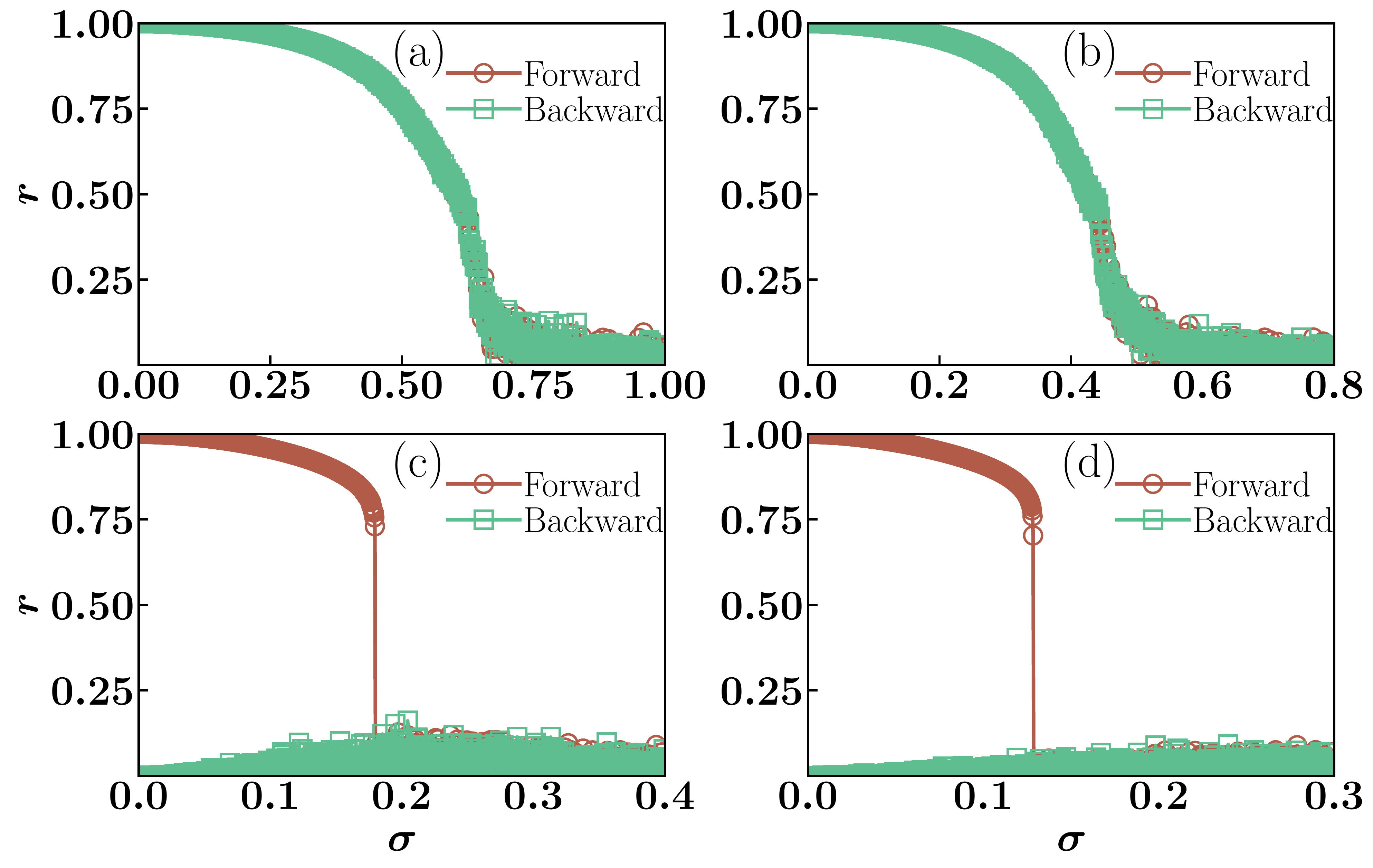}
\caption{(Color online) Variation of order parameter $r$ with
adiabatically-tuned $\sigma$ in the stationary state of the
dynamics~(\ref{eq:eom2}) on a
lattice of size $L=3200$ and
for four values of $K$, namely, $K=0.04$ (panel (a)), $K=-0.1$ (panel
(b)), $K=-0.21$ (panel (c)), and $K=-0.23$ (panel (d)). The frequency distribution $g(\omega)$ is a
Gaussian with zero mean and unit variance. The results correspond to a
typical realization of the frequencies. Hysteresis behaviour is observed only in panels
(c) and (d).  The data are obtained from numerical
integration of the dynamics~(\ref{eq:eom2}).} 
\label{fig:hysteresis}
\end{figure}

\begin{figure}[!ht]
\hspace{-0.52 cm}
\includegraphics[scale=0.4]{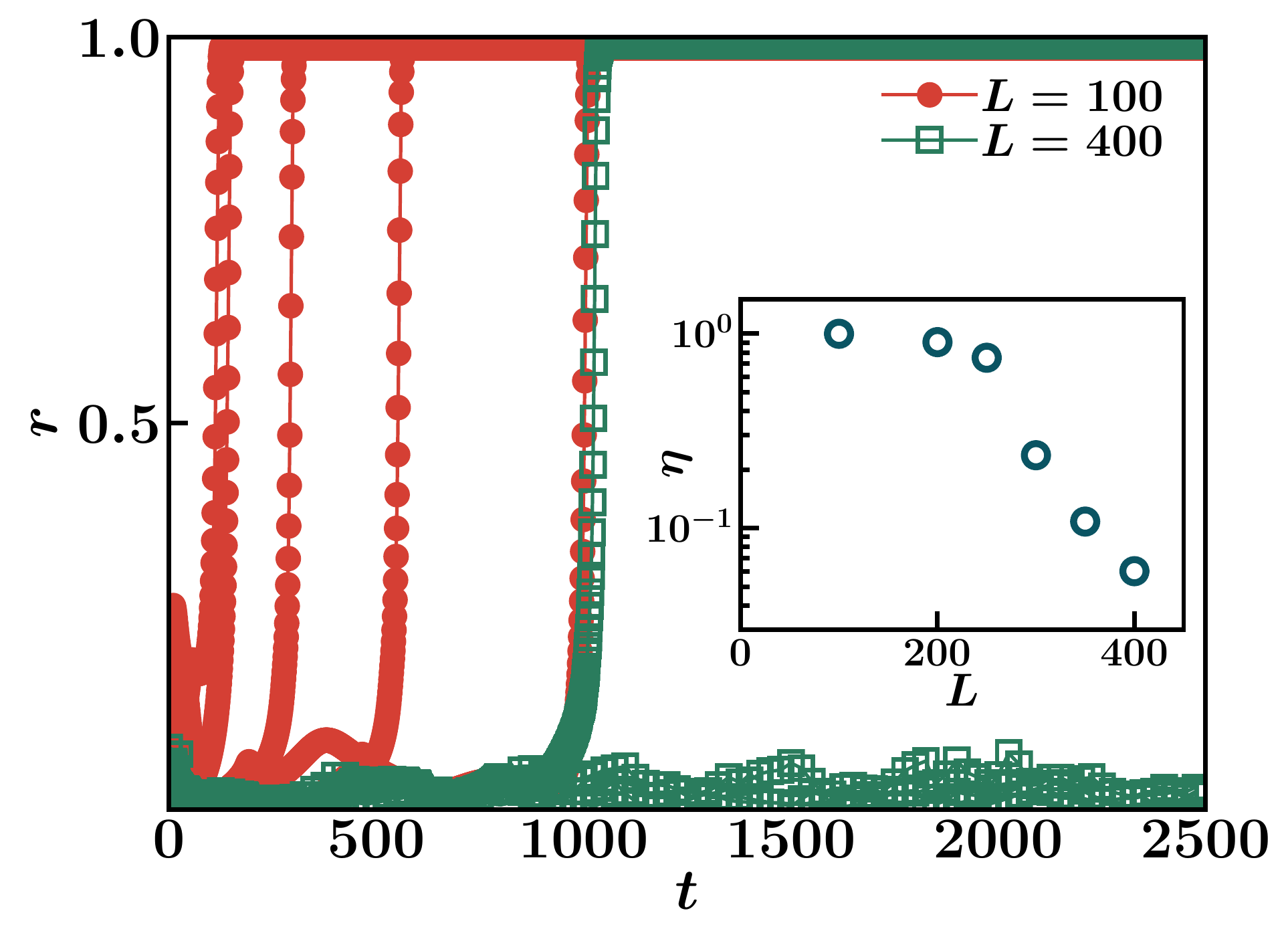}
\caption{(Color online) Considering the dynamics~(\ref{eq:eom2}) for a large 
negative value of $K$ (namely, $K=-0.21$) and at a fixed $\sigma$ at which an initial
synchronized state is stable (we have taken $\sigma=0.05$), the main figure shows for two system sizes
the time evolution of $r$ for five
realizations of an initial incoherent state. The frequency distribution $g(\omega)$ is a
Gaussian with zero mean and unit variance. It may be seen that with
increase of $L$, a smaller number of initial realizations relax to the
synchronized state over the time window of
observation. The inset shows this fraction $\eta$ as a function of $L$,
indicating fast decrease with increase of $L$. This result implies that
in the limit of large $L$, the system does not exhibit relaxation to the
synchronized state but remains close to the initial incoherent state, consistent with the results displayed in
Fig.~\ref{fig:hysteresis}, panels (c) and (d). The data are obtained from numerical
integration of the dynamics~(\ref{eq:eom2}).}
\label{fig:metastability}
\end{figure}

Since a clear distinguishing feature between first-order and continuous
transitions is the occurrence of hysteresis in the
former~\cite{Binder1987}, we now
proceed to report on results of such a study. Numerical results reported
in Fig.~\ref{fig:hysteresis} correspond to the situation in which for a
fixed value of $\sigma$, we let
the system relax to the stationary state at $\sigma=0$ while starting
from an initial state in which all the oscillators have the same angle, and then tune
$\sigma$ adiabatically to high values and back in a cycle, while
recording concomitantly the value of the order parameter $r$. Adiabatic
tuning ensures that the system is at every instant of time close to a
stationary state as $\sigma$ is tuned in time.
Figures~\ref{fig:hysteresis}(a),(b) show the variation of $r$ with
adiabatically-tuned $\sigma$, for $K=0.04$ and
$K=-0.1$, respectively. In both cases, the curves corresponding to
forward and backward variation of $\sigma$ coincide up to numerical
precision, and consequently, we do not observe any hysteresis behavior,
thereby hinting at the corresponding transition from the synchronized to
the incoherent phase being a continuous one. On the other hand, results
displayed in Figs.~\ref{fig:hysteresis}(c),(d) for $K=-0.21$ and $-0.23$,
respectively, show the existence of a hysteresis loop, thereby bearing a
clear signature of a first-order transition. It may be noted from the
results for the backward variation of $\sigma$ shown in panels (c) and
(d) that $r$ does not attain the value of unity as $\sigma$ is reduced
to zero, but instead has a value close to zero. We understand this as
due to the system being stuck in long-lived metastable states during relaxation to a
synchronized state for $K$ negative and large in
magnitude. To illustrate this point, consider the plots in
Fig.~\ref{fig:metastability} for a large 
negative value of $K$ and at a fixed $\sigma$ at which an initial
synchronized state is stable. The figure shows time evolution of $r$ for several
realizations of an initial incoherent state. It may be seen that only a fraction $\eta$
of these realizations relax to the synchronized state over the time window of
observation, with the fraction decreasing fast with the increase of system size
$L$ (inset of Fig.~\ref{fig:metastability}). This result implies that in the limit of large $L$, the system does not exhibit relaxation to the
synchronized state but remains close to the initial incoherent state, consistent with the results displayed in
Fig.~\ref{fig:hysteresis}, panels (c) and (d).

\begin{figure}[!ht]
\hspace{-0.52 cm}
\includegraphics[scale=0.22]{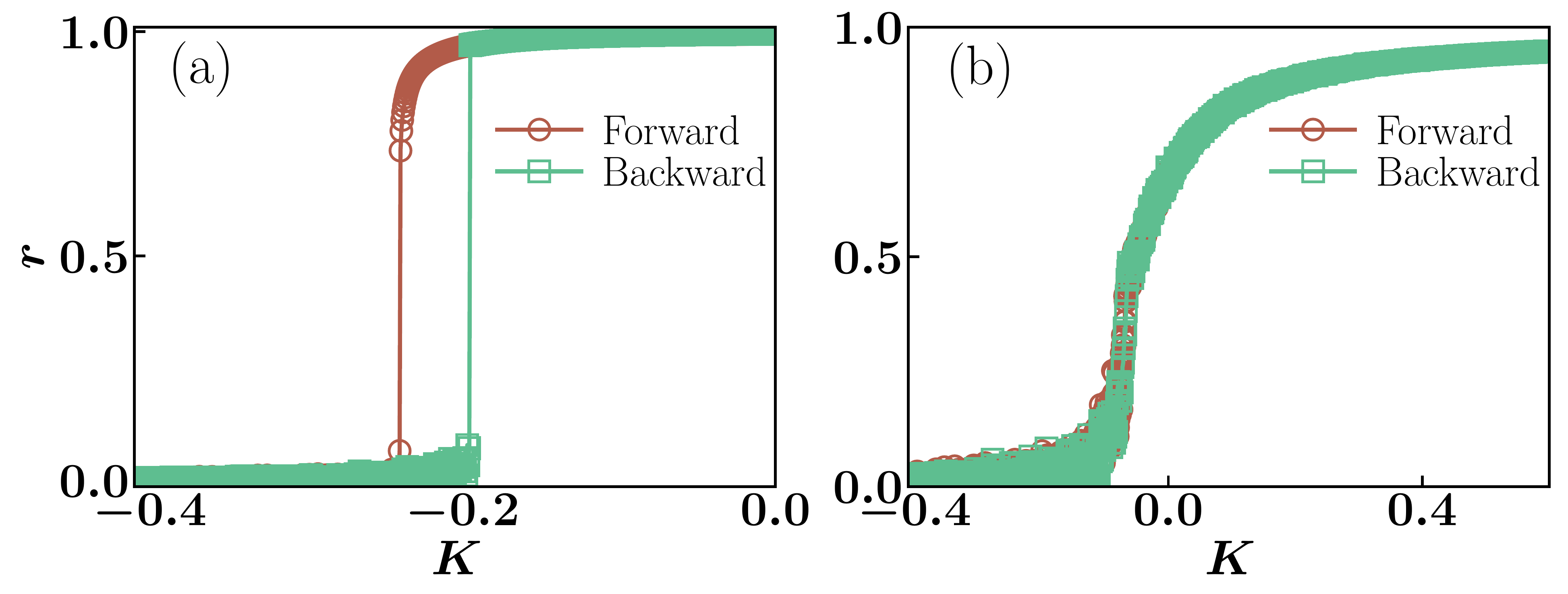}
\caption{(Color online) Variation of order parameter $r$ with
adiabatically-tuned $K$ in the stationary state of the
dynamics~(\ref{eq:eom2}) on a
lattice of size $L=3200$ and
for two values of $\sigma$, namely, $\sigma=0.1$ (panel (a)), and
$\sigma=0.5$ (panel (b)). The frequency distribution $g(\omega)$ is a
Gaussian with zero mean and unit variance. The results correspond to a
typical realization of the frequencies. Hysteresis behaviour is observed
only in panel (a) and not in panel (b), consistent with the fact that
for $\sigma=0.1$ (respectively, $\sigma=0.5$), one has a first-order
(respectively, a continuous) transition, see
Fig.~\ref{fig:parameter_space}. The data are obtained from numerical
integration of the dynamics~(\ref{eq:eom2}).} 
\label{fig:hysteresis-again}
\end{figure}

\begin{figure}[!ht]
\hspace{-0.52 cm}
\includegraphics[scale=0.22]{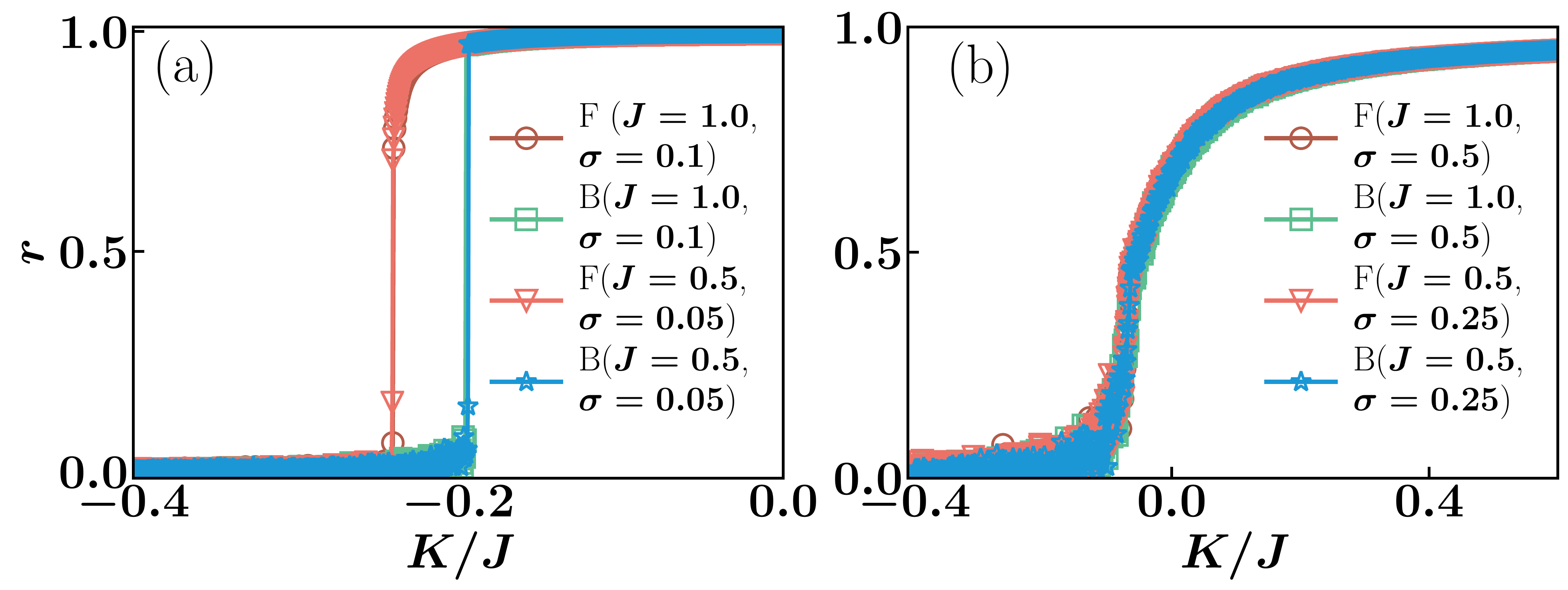}
\caption{(Color online) Variation of order parameter $r$ with
adiabatically-tuned $K$, rescaled by $J$, in the stationary state of the
dynamics~(\ref{eq:eom1-1}) on a
lattice of size $L=3200$. The values of $J$ and $\sigma$ are mentioned
in the individual panels for both forward (F) and backward (B) variation
of $K$ in time. The frequency distribution $g(\omega)$ is a
Gaussian with zero mean and unit variance. The results correspond to a
typical realization of the frequencies. The data are obtained from numerical
integration of the dynamics~(\ref{eq:eom1-1}). The results are a clear
demonstration of the fact that the relevant parameters to obtain phase transitions for the
model~(\ref{eq:eom1-1}) are the ratios $\sigma/J$ and $K/J$, as a result of which $r$ when plotted as a
function of $K/J$ and for a fixed $\sigma/J$, with different values of $J
\ne 0$, all coincide. This justifies the transformations invoked in reducing the dynamics~(\ref{eq:eom1-1})
to~(\ref{eq:eom2}).}
\label{fig:hysteresis-again-1}
\end{figure}

{ Figure~\ref{fig:hysteresis-again} shows the variation of the order parameter $r$ with
adiabatically-tuned $K$ in the stationary state of the
dynamics~(\ref{eq:eom2}) for two values of $\sigma$, namely, $\sigma=0.1$ (panel (a)), and
$\sigma=0.5$ (panel (b)). Hysteresis behaviour is observed
only in panel (a) and not in panel (b), consistent with the fact that
for $\sigma=0.1$ (respectively, $\sigma=0.5$), one has a first-order
(respectively, a continuous) transition, see
Fig.~\ref{fig:parameter_space}. As claimed following
Eq.~(\ref{eq:eom2}), Fig.~\ref{fig:hysteresis-again-1} demonstrates that
the relevant parameters to obtain our observed phase transitions for the
model~(\ref{eq:eom1-1}) are the
ratios $\sigma/J$ and $K/J$, as a result of which $r$ when plotted as a
function of $K/J$ and for a fixed $\sigma/J$, with different values of $J
\ne 0$, all coincide. This justifies the transformations invoked in reducing the dynamics~(\ref{eq:eom1-1}) to~(\ref{eq:eom2}).

One may wonder as to why the plots in Fig.~\ref{fig:rho_vs_sigma} corresponding to first-order
transitions do not show hysteresis, while the ones in
Figs.~\ref{fig:hysteresis} and~\ref{fig:hysteresis-again} do show
hysteresis. To understand this, attention may be called to the fact the
plots in Fig.~\ref{fig:rho_vs_sigma} do not correspond to adiabatic tuning of the parameter
plotted in the $x$-axis. For example, the plots of $r$ versus $K$ at a
given value of $\sigma$ correspond of several independent numerical runs
at the given $\sigma$ for each of which $K$ is fixed at given values,
letting each
run relaxing the system to stationarity and
recording the corresponding stationary value of $r$. In contrast, the plots in, e.g.,
Fig.~\ref{fig:hysteresis-again} correspond to a
single numerical run in which in the stationary state and for a fixed $\sigma$, the parameter $K$ is
continuously and adiabatically
tuned in time and the corresponding value of $r$ is recorded. As follows
from the theory of first-order phase transitions~\cite{Binder1987}, it is only in the
latter case of adiabatic tuning that one should observe hysteresis and
not in the case of Fig.~\ref{fig:rho_vs_sigma}.}

\begin{figure*}[!ht]
\centering
\includegraphics[scale=0.4]{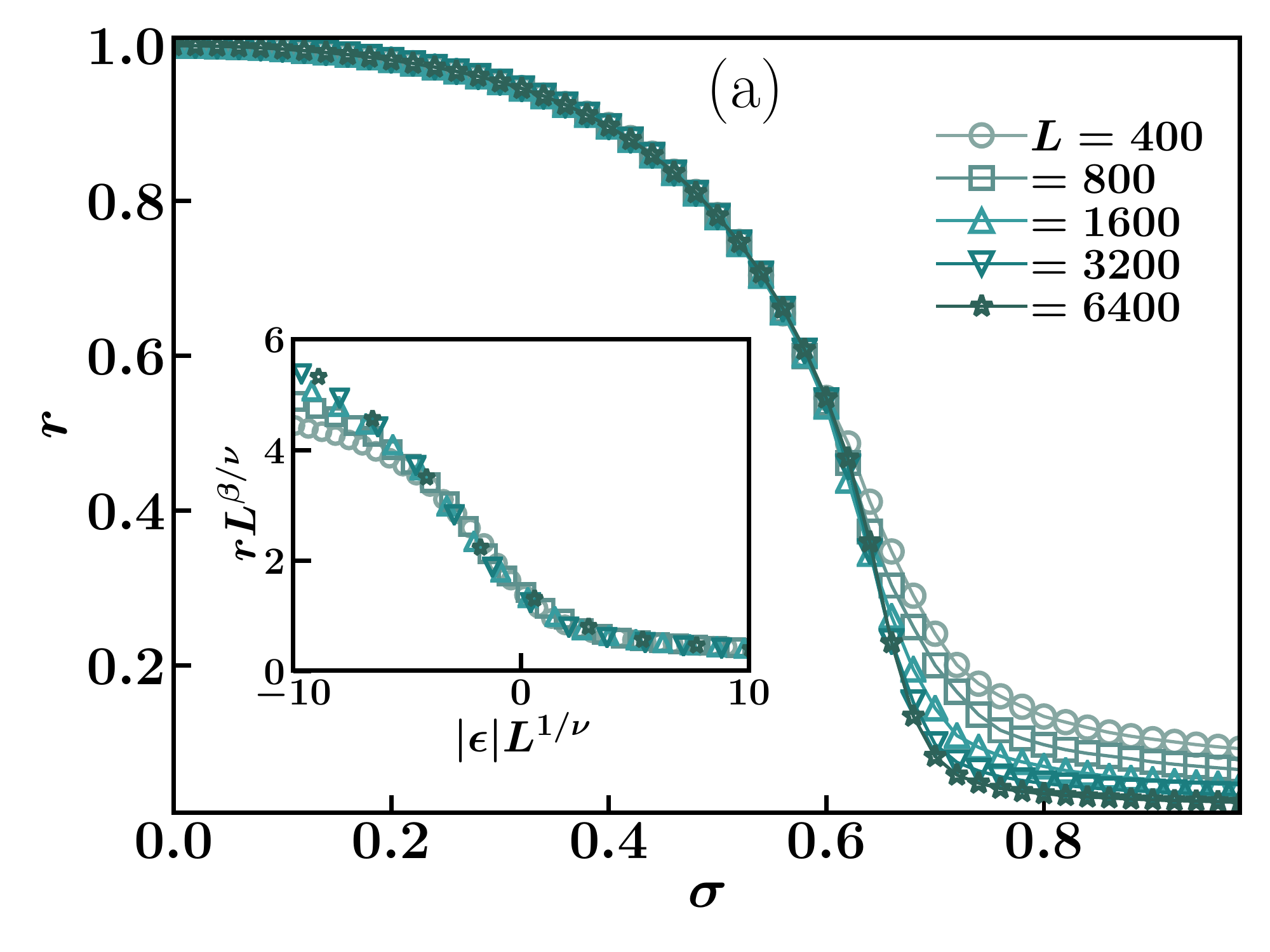}
\includegraphics[scale=0.4]{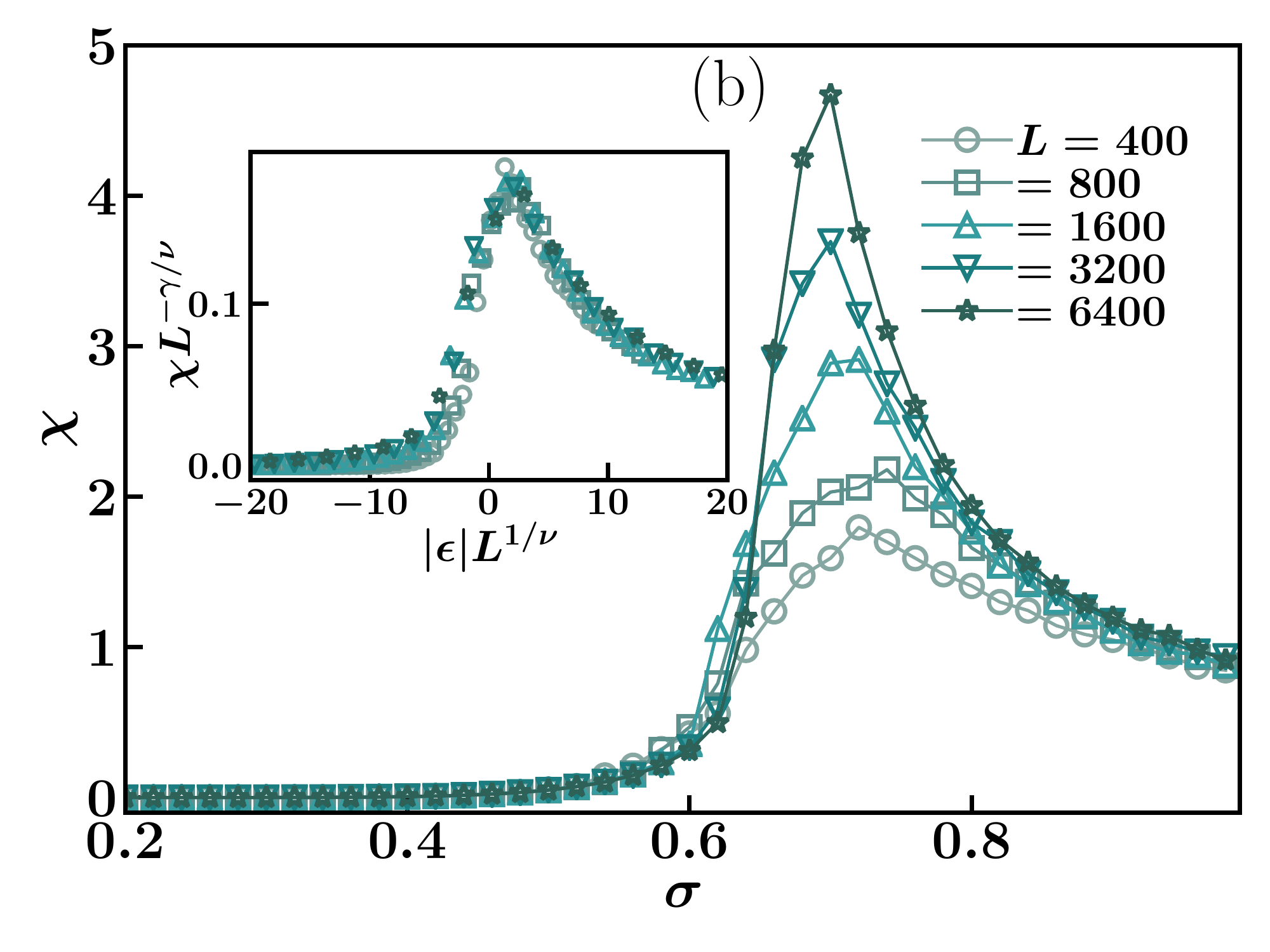}
\includegraphics[scale=0.4]{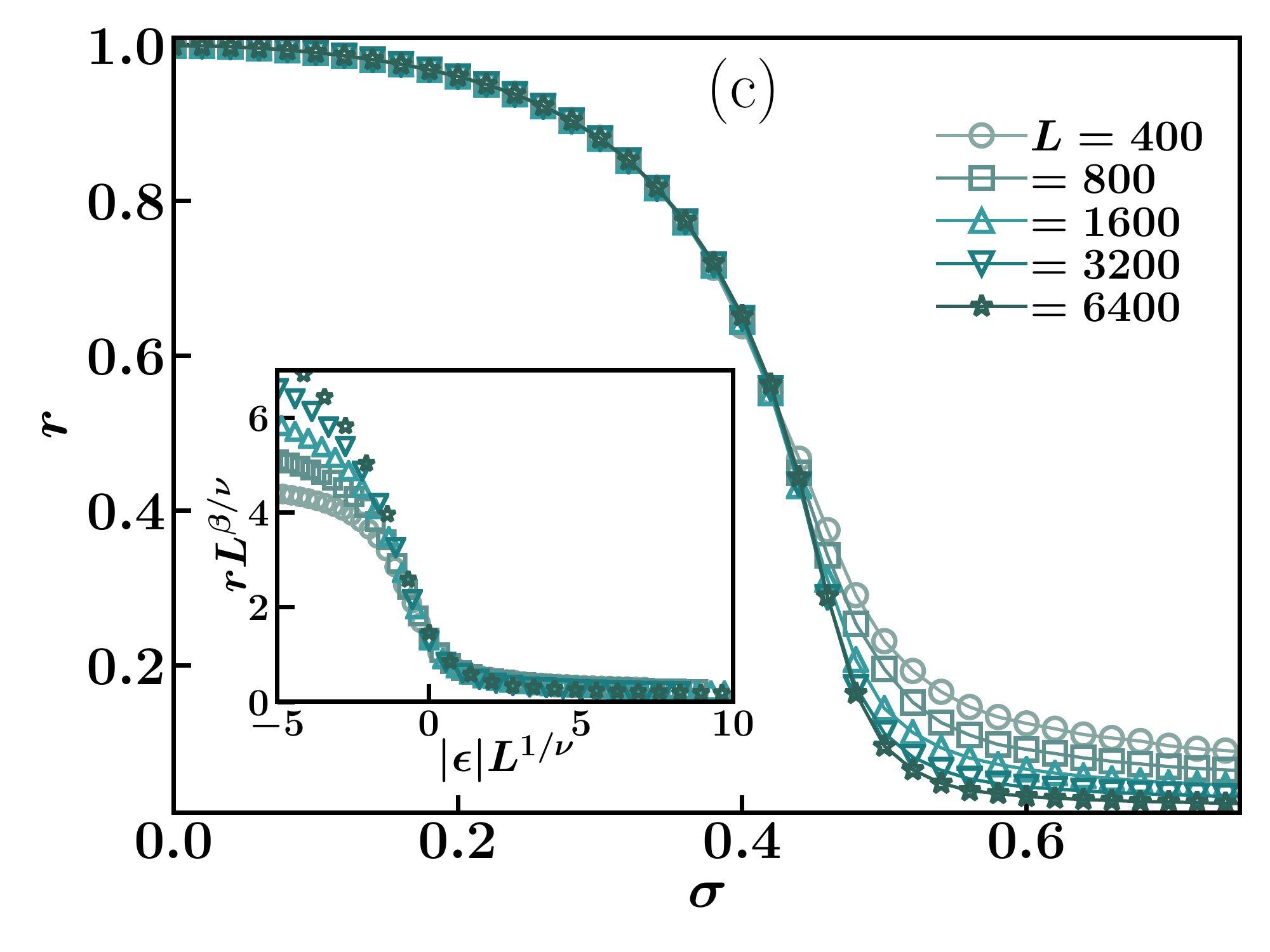}
\includegraphics[scale=0.4]{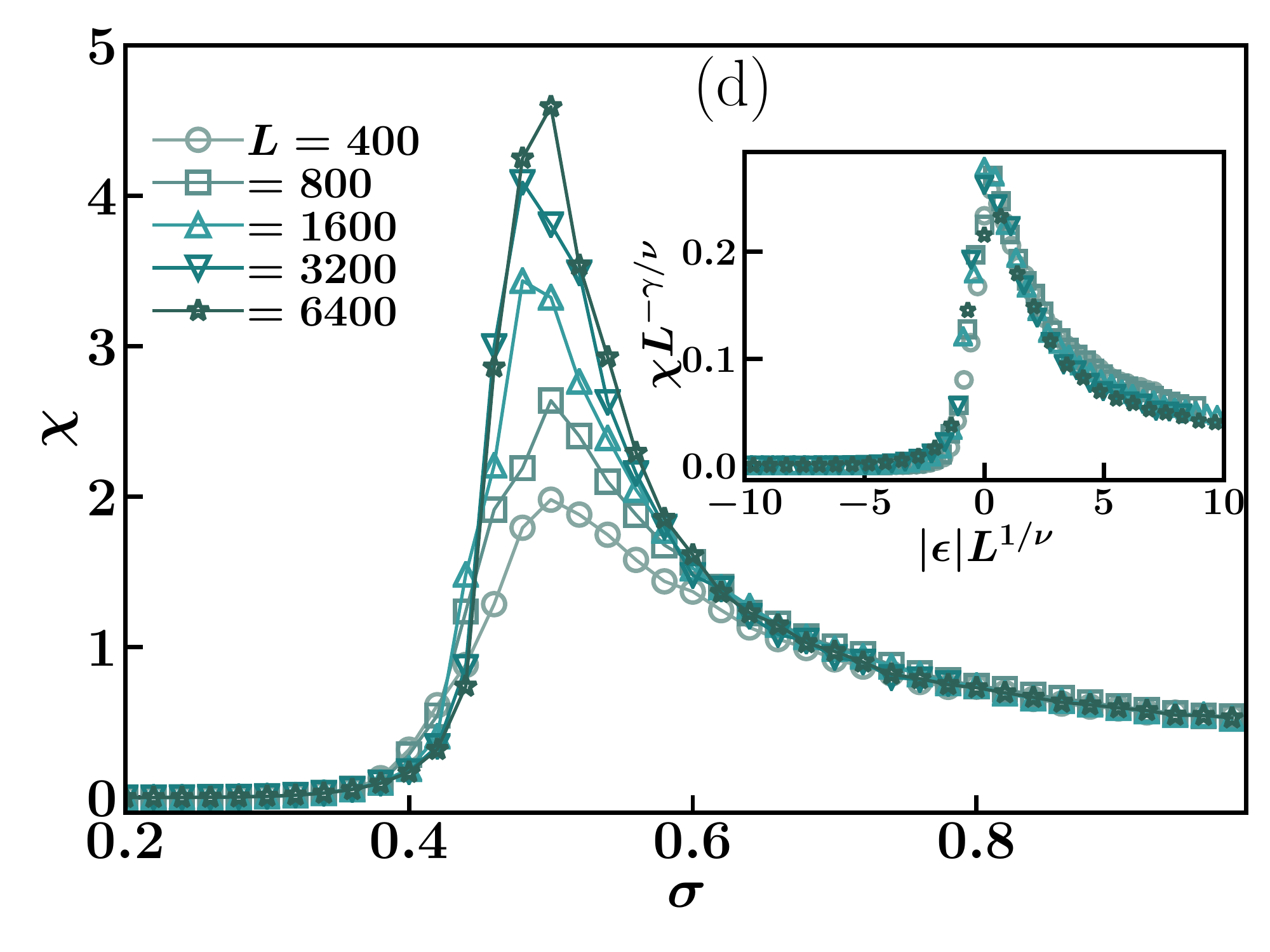}
\caption{(Color online) For two values of $K$, the figure shows the
variation with $\sigma$ of the finite-$L$ order parameter $r \equiv
r(L)$ and the quantity 
$\chi \equiv \chi(L)$ (see Eq.~(\ref{eq:chi-definition})) in the
stationary state of the model~(\ref{eq:eom2}), for five values of the system size $L$. In the insets, we
show scaling collapse of the data according to
Eq.~(\ref{eq:scaling-form}). The frequency distribution $g(\omega)$ is a
Gaussian with zero mean and unit variance. We have $K=0.04$ for panels (a) and (b) and
$K=-0.1$ for panels (c) and (d). The data involve time averaging in the stationary state
for a given frequency realization $\{\omega_i\}$ as well as over
different frequency realizations. The critical point $\sigma_c \equiv
\sigma_c(K)$ is obtained by plotting the maximum of $\chi(L)$ as a
function of $L$ and fitting it to a power law, while the values of the critical exponents
$\beta,\nu,\gamma$ are obtained from the
scaling collapse of the data for $r$ and $\chi$. The data are obtained from numerical
integration of the dynamics~(\ref{eq:eom2}).}
\label{fig:fss-second-order}
\end{figure*}

On the basis of the foregoing, we may conclude the existence of both
continuous and first-order phase transitions in the stationary state of
the dynamics~(\ref{eq:eom2}). Our next task would be to explain how we obtain numerically  the
phase-transition lines in the $(\sigma,K)$-plane, as shown in
Fig.~\ref{fig:parameter_space}, and to explain in particular how
we locate the tricritical point, defined as the point at which the first-order and continuous
transition lines meet. In the following, we will discuss the phase
diagram and the phase transitions presented therein by varying $\sigma$
at a fixed $K$, though we emphasize that this way of uncovering the
nature of the phase transition is completely equivalent to varying $K$
at a fixed $\sigma$, with the latter being perhaps more amenable to experimental
implementation; the equivalence is evidently true from the
phase diagram~\ref{fig:parameter_space}.

\subsection{Obtaining the line of continuous transition}
\label{sec:continuous}

In order to locate numerically the line of continuous transition, we proceed as
follows. At values of $K$ at which no hysteresis is observed in the
variation of $r$ with adiabatically-tuned $\sigma$, our aim
is to estimate the value of $\sigma_c \equiv \sigma_c(K)$, namely, the value of $\sigma$ at the critical point of
transition at fixed $K$. To this end, we analyze the
finite-$L$ data for stationary $r$ by resorting to the finite-size scaling theory for equilibrium critical
phenomena briefly summarized in Appendix \ref{app1}. By drawing an analogy with
Eq.~(\ref{eq:app-scaling-form}), we write scaling forms for the order
parameter $r(L)$ obtained in a system of size $L$ and the
stationary-state temporal fluctuations of the order parameter defined as
\begin{equation}
\chi(L) \equiv L~\overline{\langle r^2(L)\rangle - \langle r(L)
\rangle^2},
\label{eq:chi-definition}
\end{equation}
where the angular brackets and the overbar denote respectively time average
in the stationary state for a given frequency realization $\{\omega_i\}$ and  average over frequency realizations.
The scaling forms are 
\begin{eqnarray}
&&r(L) \sim L^{-\beta/\nu} f(|\epsilon|L^{1/\nu}), \nonumber \\
\label{eq:scaling-form} \\
&&\chi(L) \sim L^{\gamma/\nu}g(|\epsilon| L^{1/\nu}), \nonumber 
\end{eqnarray}
with $\beta,\nu,\gamma$ being the critical exponents, and 
\begin{equation}
\epsilon \equiv \frac{\sigma-\sigma_c}{\sigma_c}.
\end{equation}
As discussed in Appendix \ref{app1}, the scaling functions $f(x)$ and $g(x)$, defined with $x>0$,
behave in the limit $x\to \infty$ as $f(x) \sim x^\beta$ and $g(x)\sim
x^{-\gamma}$. In the limit $x \to 0$, both the functions behave as
constants.

Now, following the procedure detailed in Appendix \ref{app1} to obtain
the critical point, $\sigma_c \equiv
\sigma_c(K)$ is estimated from the plot of the maximum of $\chi(L)$ as a
function of $L$ and fitting it to a power law. Using the value of $\sigma_c$ estimated this way,
and requiring for large $L$ scaling collapse of the finite-$L$ data for
$r(L)$ and $\chi(L)$ according to the
forms in Eq.~(\ref{eq:scaling-form}) allow to obtain values for the critical exponents $\beta,
\gamma, \nu$. In Fig.~\ref{fig:fss-second-order}, we show for two values
of $K$ the behavior of $r$ (panels (a) and (c))  and $\chi$ (panels (b)
and (d)) as a function of $\sigma$ and scaling collapse in the
corresponding insets. We have $K=0.04$ for panels (a) and (b) and
$K=-0.1$ for panels (c) and (d). The values of the
critical exponents that yielded scaling collapse are: for $K=0.04$, we
have $\beta \approx 0.52, \nu \approx 2.0, \gamma \approx 0.76$, while for $K=-0.1$, we have
$\beta \approx 0.78,\nu \approx 3.13,\gamma \approx 1.06$. We note that one requires
data for larger $L$ in order to estimate more reliably the critical exponent
values. Our focus here is primarily on establishing the existence of a continuous phase transition in the dynamics~(\ref{eq:eom2}) for a range of
values of $K$, and in this regard, a confirmation, in addition to
the no-hysteresis data presented in Fig.~\ref{fig:hysteresis}, is
provided by the very good scaling collapse for large $L$ demonstrated in
Fig.~\ref{fig:fss-second-order} for which the underlying theory invoked
is that of finite-size scaling for continuous transitions. That we have
been able to estimate $\sigma_c$ accurately is evident from the quality
of scaling collapse seen in Fig.~\ref{fig:fss-second-order}. 

The aforementioned
procedure of obtaining $\sigma_c(K)$ from the data of $\chi(L)$ is
repeated for several values of $K$ at which one
does not observe any hysteresis in the behavior of $r$ as a function of
adiabatically-tuned $\sigma$. In this way, we obtain the values of $\sigma_c(K)$ as a
function of $K$, which we use to construct the phase diagram in the
$(\sigma,K)$-plane, that is, draw the line of continuous transition, see Fig.~\ref{fig:parameter_space}.

\subsection{Obtaining the line of first-order transition}
\label{sec:first-order}

\begin{figure}[ht]
\hspace{-0.8 cm}
\includegraphics[scale=0.45]{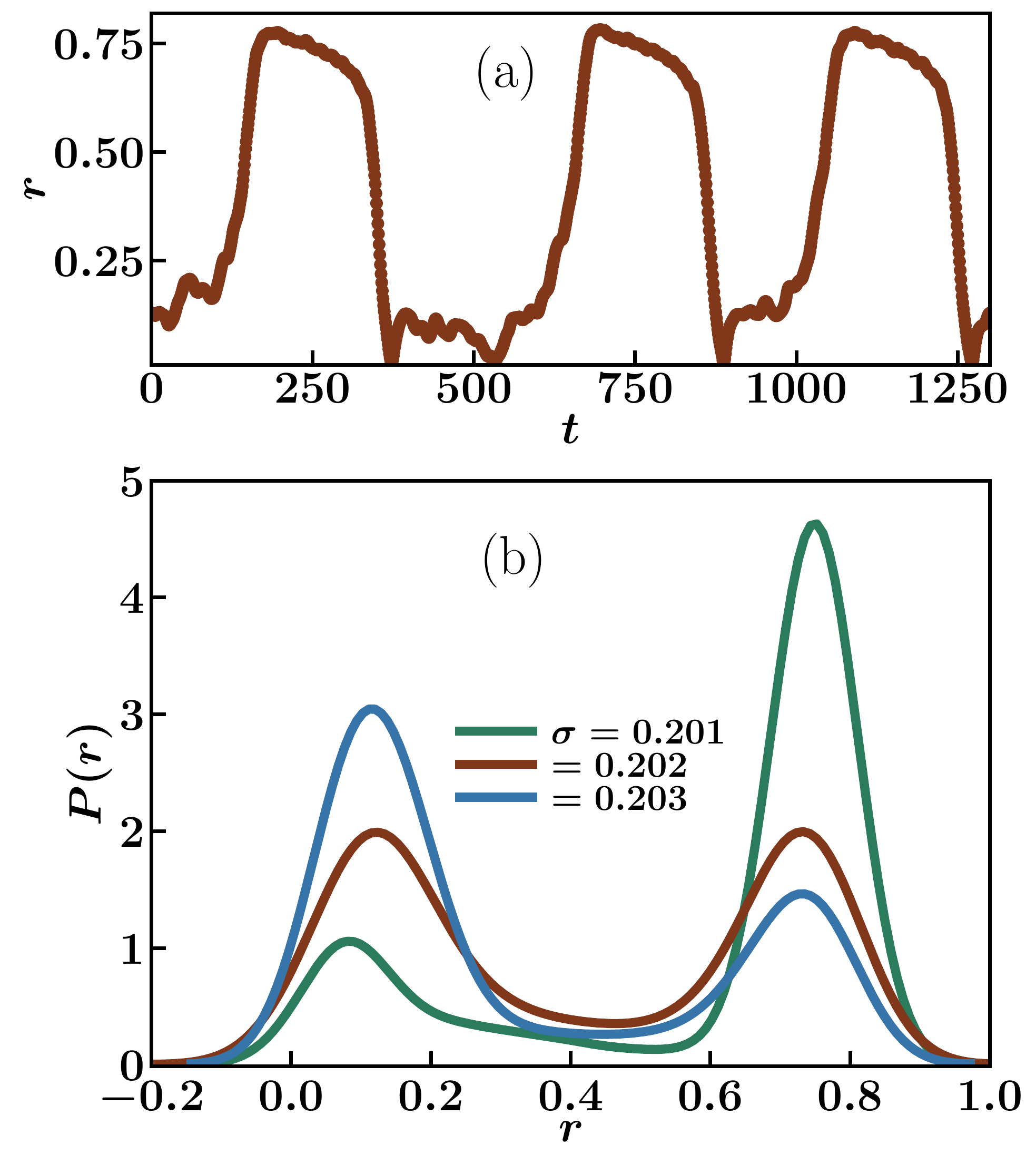}
\caption{(Color online) Time variation of the order parameter
$r$ (panel (a)) and the corresponding distribution $P(r)$ (panel (b)) in
the stationary state of the dynamics~(\ref{eq:eom2}) and at a value of
$K$ (namely, $K=-0.202$) at which one has a first-order transition. The system size is
$L=3200$. The frequency distribution $g(\omega)$ is a
Gaussian with zero mean and unit variance. The results correspond to a
typical realization of the frequencies. Exactly at the transition
point, the distribution has two peaks of equal height, while on either
side, the peaks have different heights. Note that the switching time between the two bistable states, as in
panel (a), becomes with increasing $L$ so prohibitively large that one
does not observe enough switching within a reasonable time interval of
observation, and then, one does not have enough statistics to draw the
distribution $P(r)$. The data are obtained from numerical
integration of the dynamics~(\ref{eq:eom2}).}
\label{fig:bistability}
\end{figure}

Having obtained in the preceding section the line of continuous
transition, we now proceed to obtain the line of first-order transition.
In the absence of a scaling theory akin to the one that exists on
general grounds for continuous transitions, we proceed to obtain the
first-order transition point as follows. At a first-order phase
transition, the order parameter as a function of time shows bistability,
with the system switching back and forth between two phases. For
our system~(\ref{eq:eom2}), we show in Fig.~\ref{fig:bistability}(a) the
behavior of $r$ versus time in the stationary state and at a value of
$K$ at which we have observed hysteresis (\textit{cf.}
Fig.~\ref{fig:hysteresis}). Such a bistable behavior may be characterized
by drawing the probability distribution $P(r)$ of stationary $r$. When
bistable, $P(r)$ is bimodal with two peaks of equal heights. Contrarily, while on
either side of the transition point when the system is no more bistable,
the distribution $P(r)$ is bimodal, but the peaks are not of equal
heights. Considering our model~(\ref{eq:eom2}), when one is at a value
of $\sigma$ smaller (respectively, greater) than the critical value of
first-order transition, $P(r)$ will have a higher peak at a value of $r$
corresponding to the synchronized (respectively, incoherent) phase.
Then, in order to locate the transition point, we adopt the following
strategy. For
a fixed $K$ and a given (large) system size $L$, we scan the range of $\sigma$,
obtaining for each value the distribution $P(r)$ from the time variation
of $r$ in the stationary state, and estimate the transition point as the value of $\sigma$ at
which $P(r)$ has two peaks of equal heights. An example is shown in
Fig.~\ref{fig:bistability}(b). Note that unlike a
first-order transition point that is characterized by two equally likely
values of the order parameter, a continuous transition is characterized
by a distribution $P(r)$ that is single peaked, with the peak shifting
continuously from non-zero to zero values as $\sigma$ is tuned from below to above
the transition point.

The above background on how to locate first-order and
continuous transition points in the $(\sigma,K)$-plane armed us to draw
in Fig.~\ref{fig:parameter_space} the corresponding transition lines and
to locate the tricritical point at which the two lines meet.

In the following section, we embark on an analysis of the
dynamics~(\ref{eq:eom2}) based on an approximate theory that allows to
obtain the behavioral trend of the order parameter in the stationary state.

\section{Theoretical analysis}
\label{sec:tat}

In this section, we discuss a suitably-modified version of an
approximate time-averaged theory proposed in~\cite{restrepo2005onset},
see also~\cite{Rodrigues2016}, which allows to obtain quite accurately the behavior of the
order parameter in the stationary
state of our model~(\ref{eq:eom2}) in parameter regimes of continuous transitions. To proceed, let us define a weighted
adjacency matrix as  
\begin{equation}
W_{ij}\equiv\frac{1}{L}(1-\delta_{ij})+ K \delta_{i,j\pm 1};~~i,j=1,2,\ldots, L,
\end{equation}
in terms of which we rewrite
Eq.~(\ref{eq:eom2}) as 
\begin{equation}
\frac{{\rm d} \theta_{i}}{{\rm d}t}=\sigma \omega_{i} + {\rm Im}\left[e^{-i\theta_i}\sum_{j=1}^L
W_{ij}e^{i\theta_j}\right].
\end{equation}
Let us now consider the above dynamics in the stationary state, and express it as 
\begin{eqnarray}
\frac{{\rm d} \theta_{i}}{{\rm d}t}&=& \sigma \omega_{i} + r_i^{({\cal T})}\sin(\psi_i-\theta_i)+h_i(t),
\label{eq:tat1}
\end{eqnarray}
where we have defined a time-averaged local order parameter for the $i$-th site as
\begin{eqnarray}
&&r_{i}^{({\cal T})} e^{i \psi_i} \equiv \sum_{j=1}^{L} W_{ij} \langle e^{i
\theta_{j}}\rangle\nonumber \\
&&=re^{i\psi}+K\left(\langle
e^{i\theta_{i+1}}\rangle + \langle
e^{i\theta_{i-1}}\rangle\right)-\frac{1}{L}\langle e^{i\theta_i}\rangle,
\label{eq:rT-definition}
\end{eqnarray}
with the angular brackets denoting as usual time average over dynamics in the
stationary state for a given frequency realization $\{\omega_i\}$, while $h_i(t)$ denotes stationary-state fluctuations:
\begin{equation}
h_i(t) \equiv 
{\rm Im}\left[e^{-i\theta_i}\sum_{j=1}^L
W_{ij}\left(e^{i\theta_j}-\langle e^{i\theta_j}\rangle\right)\right].
\label{eq:hi}
\end{equation}
In obtaining the first term on the rhs of Eq.~(\ref{eq:rT-definition}), we have used the
fact that since we are in the stationary state, we have $\langle
re^{i\psi}\rangle=re^{i\psi}$. Note that the quantities $r_i^{({\cal
T})}$ and $\psi_i$ are by definition time
independent.

The time-averaged theory aims to study the synchronized phase by
neglecting for large $L$ the fluctuations
$h_i(t)$ in the dynamics~(\ref{eq:tat1})~\cite{restrepo2005onset,Rodrigues2016}, which therefore reads
\begin{equation}
\frac{{\rm d} \theta_{i}}{{\rm d}t}=\sigma \omega_{i} + r_i^{({\cal T})}\sin(\psi_i-\theta_i).
\label{eq:tat2}
\end{equation}
Considering the dynamics~(\ref{eq:tat2}), it is well known from the study of a similar equation occurring in the
usual Kuramoto model~\cite{strogatz2000,Gupta2018} that if the $i$-th
oscillator has $r_i^{({\cal T})}$ having such a value that $\sigma|\omega_i|\le
r_i^{({\cal T})}$, the quantity $\theta_i-\psi_i$ would have a stable fixed point
given by $\sin(\theta_i-\psi_i)=\sigma\omega_i/r_i^{({\cal
T})};~\cos(\theta_i-\psi_i)=+\sqrt{1-\sigma^2\omega_i^2/(r_i^{({\cal
T})})^2}$, the latter determining the value of $\theta_i-\psi_i$ in the
stationary state. All such oscillators satisfying $\sigma|\omega_i|\le
r_i^{({\cal T})}$ are therefore called phase-locked or
synchronized oscillators. On the other hand, oscillators with
$\sigma|\omega_i|>r_i^{({\cal T})}$ constitute the so-called drifting
oscillators, for which the dynamics~(\ref{eq:tat2}) does not allow for a stable fixed point. 

Let $\rho_j(\theta)d\theta$ denote the stationary probability that the $j$-th
oscillator, with its natural frequency equal to $\omega_j$, has its
angle in the range ($\theta, \theta+{\rm d}\theta$). If the $j$-th
oscillator is phase locked, the normalized density is given
by~\cite{acebron2005kuramoto,Gupta2018} 
\begin{eqnarray}
\rho_j^{\rm locked}(\theta-\psi)&=&r_j^{({\cal T})}\cos (\theta-\psi)\delta\left(\sigma\omega_j-r_j^{({\cal
T})}\sin(\theta-\psi)\right)\nonumber \\
&&\times\Theta\left(\cos (\theta-\psi)\right),
\label{eq:locked_prob_density}
\end{eqnarray}
with $\Theta(x)$ being the Heaviside step function. On the other hand,
the probability density in the case that the $j$-th oscillator is
drifting is given by~\cite{acebron2005kuramoto,Gupta2018}
\begin{equation}
\rho_{j}^{\rm
drift}(\theta-\psi)=\frac{1}{2\pi}\frac{\sqrt{\sigma^{2}{\omega_j}^2-(r_j^{({\cal
T})})^2 }}{|\sigma \omega_{j} - r_j^{({\cal T})}\sin(\theta-\psi)|}.
\label{eq:drift_prob_density}
\end{equation}

The value of $r_i^{({\cal T})}$ may then be found self-consistently as  
\begin{eqnarray}
&&r_i^{({\cal T})}= r_i^{(\cal T)}\big\vert_{\rm locked} + r_i^{(\cal T)}\big \vert_{\rm drift} \nonumber\\
&&=\sum_{j;~\sigma|\omega_j| \le r_j^{({\cal
T})}}W_{ij}\langle
e^{i(\theta_j-\psi_i)}\rangle +\sum_{j;~\sigma|\omega_j|>r_j^{({\cal
T})}}W_{ij}\langle
e^{i(\theta_j-\psi_i)}\rangle. \nonumber \\\label{eq:rsync} 
\end{eqnarray}
The contribution of the locked oscillators to the order parameter is calculated as follows:
\begin{equation}
r_i^{({\cal T})}\big\vert_{\rm locked}=\sum_{j;~\sigma|\omega_j| \le
r_j^{({\cal T})}}W_{ij}\langle
e^{i(\theta_j-\psi_j)}e^{i(\psi_j-\psi_i)}\rangle.
\label{eq:rsync-1}
\end{equation}
These oscillators have $\theta_j-\psi_j$ taking up time-independent
values in
the stationary state, so that the corresponding factor may be taken out of the angular
brackets in Eq.~(\ref{eq:rsync-1}), Moreover, $\psi_i$ and $\psi_j$
being time independent, we have $\langle
e^{i(\psi_j-\psi_i)}\rangle=e^{i(\psi_j-\psi_i)}$. The time-independent
values for $\theta_j-\psi_j$ are distributed according to the
delta-function distribution~(\ref{eq:locked_prob_density}), implying
that we have $(\theta_j-\psi_j)=\sin^{-1}\left(\sigma\omega_j/r_j^{({\cal
T})}\right);~\cos(\theta_j-\psi_j)=+\sqrt{1-\sigma^2\omega_j^2/(r_j^{({\cal
T})})^2}$). Putting all these together, we have
\begin{eqnarray}
r_i^{({\cal T})}\big\vert_{\rm
locked}&=&\sum\limits_{j;~\sigma|\omega_j| \le r_j^{({\cal T})}} W_{ij}
e^{i(\psi_j-\psi_i)}\nonumber \\
&&\times\left[\left(\sqrt{1-\frac{\sigma^2\omega_j^2}{(r_j^{({\cal
T})})^2}}\right)+i\left(\frac{\sigma\omega_j}{r_j^{({\cal
T})}}\right)\right].
\label{eq:r_locked}
\end{eqnarray}

Proceeding in the same manner as for the locked oscillators, we may obtain the contribution of the drifting oscillators:
\begin{eqnarray}
&&r_i^{({\cal T})}\big\vert_{\rm drift}=\sum_{j;~\sigma|\omega_j| >
r_j^{({\cal T})}}W_{ij}\langle
e^{i(\theta_j-\psi_j)}e^{i(\psi_j-\psi_i)}\rangle \nonumber \\
&&=\sum_{j;~\sigma|\omega_j| > r_j^{({\cal T})}}W_{ij} e^{i(\psi_j-\psi_i)} \langle
e^{i(\theta_j-\psi_j)}\rangle \nonumber \\
&&=\sum_{j;~\sigma|\omega_j| > r_j^{({\cal T})}}W_{ij} e^{i(\psi_j-\psi_i)}\left[ \langle
\cos(\theta_j-\psi_j)\rangle + i \langle \sin(\theta_j-\psi_j)\rangle \right]. \nonumber \\
\label{eq:rdrift-1}
\end{eqnarray}
Now, the drifting oscillators, unlike the locked ones, do not have
time-independent values for their angle $\theta_j-\psi_j$, but instead
have their values distributed according to the stationary
distribution~(\ref{eq:drift_prob_density}). Consequently, in computing
the time average $\langle e^{i(\theta_j-\psi_j)} \rangle$, we need to
consider that $(\theta_j-\psi_j)$ would take values following the
distribution~(\ref{eq:drift_prob_density}), so that we have
\begin{eqnarray}
\langle \cos(\theta_j-\psi_j)\rangle =\int_{0}^{2\pi} {\rm
d}(\theta-\psi)~\rho_j^{\rm drift}(\theta-\psi) \cos(\theta-\psi)=0, \nonumber
\\
\label{eq:rdrift-2}
\end{eqnarray}
and
\begin{eqnarray}
\langle \sin(\theta_j-\psi_j)\rangle&=&\int_{0}^{2\pi} {\rm
d}(\theta-\psi) \rho_j^{\rm drift}(\theta-\psi) \sin(\theta-\psi) \nonumber\\
&=&\frac{\sigma \omega_j}{r_j^{({\cal
T})}}\left[1-\sqrt{1-\frac{(r_j^{({\cal T})})^2}{\sigma^2 {\omega_j}^2}}
\right],
\label{eq:rdrift-4}
\end{eqnarray}
finally yielding
\begin{eqnarray}
r_i^{({\cal T})}\big\vert_{\rm drift} &&= \sum_{j;~\sigma|\omega_j| >
r_j^{({\cal T})}}W_{ij} e^{i(\psi_j-\psi_i)} \nonumber \\
&&\times\left[ i \frac{\sigma \omega_j}{r_j^{({\cal
T})}}\left(1-\sqrt{1-\frac{(r_j^{({\cal T})})^2}{\sigma^2 {\omega_j}^2}} \right) \right]. 
\label{eq:rdrift-5}
\end{eqnarray}

Using Eqs.~(\ref{eq:r_locked}) and~(\ref{eq:rdrift-5}) in
Eq.~(\ref{eq:rsync}), and then equating real and imaginary parts
from both sides of it, we get
\begin{widetext}
\begin{eqnarray}
&&r_i^{({\cal T})}=\nonumber \\
&&\sum_{j;~\sigma|\omega_j|\le r_j^{({\cal T})}}
W_{ij}\left[\cos(\psi_j-\psi_i)\left(\sqrt{1-\frac{\sigma^2\omega_j^2}{(r_j^{({\cal
T})})^2}}\right)-\sin(\psi_j-\psi_i)\left(\frac{\sigma\omega_j}{r_j^{({\cal
T})}}\right)\right]- \sum_{j;~\sigma|\omega_j| > r_j^{({\cal T})}}
W_{ij}\left[ \sin(\psi_j-\psi_i) \frac{\sigma \omega_j}{r_j^{({\cal
T})}}\left(1-\sqrt{1-\frac{(r_j^{({\cal T})})^2}{\sigma^2 {\omega_j}^2}}
\right) \right], \nonumber \\
 \label{eq:rsync-21} \\
&&0=\nonumber \\
&&\sum_{j;~\sigma|\omega_j|\le r_j^{({\cal T})}}
W_{ij}\left[\sin(\psi_j-\psi_i)\left(\sqrt{1-\frac{\sigma^2\omega_j^2}{(r_j^{({\cal
T})})^2}}\right) +\cos(\psi_j-\psi_i)\left(\frac{\sigma\omega_j}{r_j^{({\cal
T})}}\right)\right]+ \sum_{j;~\sigma|\omega_j| > r_j^{({\cal T})}}
W_{ij}\left[ \cos(\psi_j-\psi_i) \frac{\sigma \omega_j}{r_j^{({\cal
T})}}\left(1-\sqrt{1-\frac{(r_j^{({\cal T})})^2}{\sigma^2 {\omega_j}^2}}
\right) \right]. \nonumber \\ \label{eq:rsync-22} 
\end{eqnarray}
\end{widetext}
The above equations are solved with the choice
$\psi_i=\psi_j~\forall~i,j$.
Equation~(\ref{eq:rsync-22}) then reduces to
\begin{equation}
0=\sum_{j}W_{ij}\left(\frac{\sigma\omega_j}{r_j^{({\cal T})}}\right) - 
\sum_{j;~\sigma|\omega_j| > r_j^{({\cal T})}} W_{ij} \frac{\sigma
\omega_j}{r_j^{({\cal T})}}\left(\sqrt{1-\frac{(r_j^{({\cal
T})})^2}{\sigma^2 {\omega_j}^2}} \right),
\label{eq:rsync-222}
\end{equation}
while Eq.~(\ref{eq:rsync-21}) now reads
\begin{equation}
r_i^{({\cal T})}=\sum_{j;~\sigma|\omega_j|\le r_j^{({\cal T})}}
W_{ij}\left(\sqrt{1-\frac{\sigma^2\omega_j^2}{(r_j^{({\cal
T})})^2}}\right). \label{eq:rsync-212} 
\end{equation}
Equations~(\ref{eq:rsync-222}) and~(\ref{eq:rsync-212}) are
simultaneously satisfied by taking all $r_i^{({\cal T})}$'s to be even in
$\{\omega_j\}$:~ $r_i^{({\cal T})}(\{\omega_j\})=r_i^{({\cal
T})}(\{-\omega_j\})~\forall~i$ and satisfying Eq.~(\ref{eq:rsync-212}).
With our choice of $\omega_j$'s being sampled from a
symmetric $g(\omega):~ g(\omega)=g(-\omega)$, Eq.~(\ref{eq:rsync-222})
is then automatically satisfied for large $L$, as the contributions in
the two sums for every pair of positive and negative $\omega_j$ cancel each
other. The set of $L$ coupled equations~(\ref{eq:rsync-212}) when solved numerically determines the set
$\{r_i^{({\cal
T})}\}$. Equation~(\ref{eq:rT-definition}) then allows to obtain the order
parameter $r$ for a given frequency realization $\{\omega_j\}$ as
\begin{equation}
r=\frac{1}{(1+2K)L-1}\left|\sum_{i=1}^L r_i^{({\cal
T})}e^{i\psi_i}\right|=\frac{1}{(1+2K)L-1}\left|\sum_{i=1}^L r_i^{({\cal
T})}\right|,
\label{eq:r-tat-final}
\end{equation}
where in the last step we have used the fact that all the $\psi_i$'s are
equal. Finally, we average the value of $r$ so obtained over different
frequency realizations.

\begin{figure}[]
\centering
\includegraphics[scale=0.22]{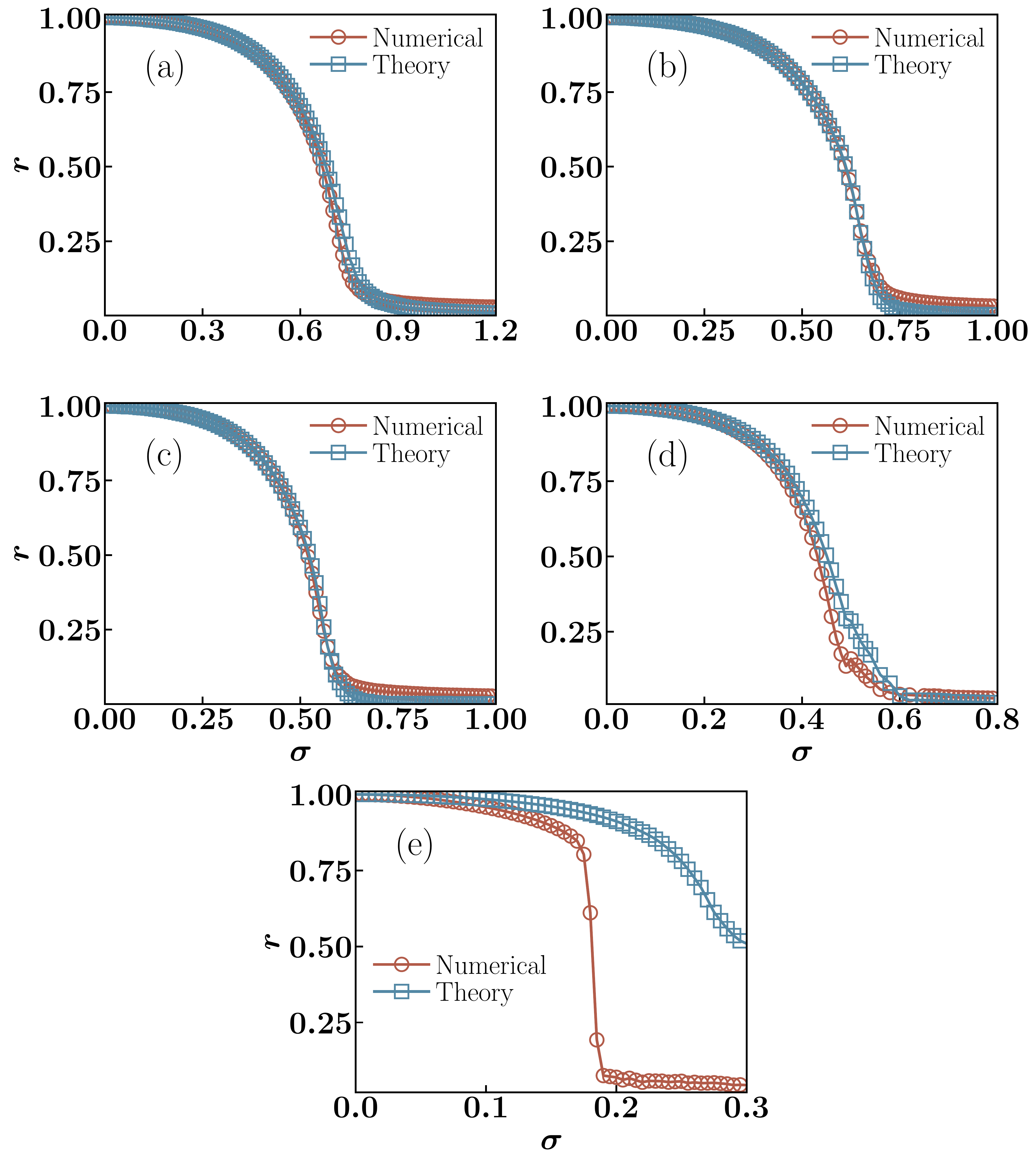}
\caption{(Color online) The figure shows the
variation with $\sigma$ of the stationary order parameter $r$ obtained
for the dynamics~(\ref{eq:eom2}) on a lattice of size $L=3200$, for five
values of $K$: $K=0.1$ (panel (a)), $K=0.04$ (panel (b)), $K=-0.04$
(panel (c)), $K=-0.1$
(panel (d)), and $K=-0.21$ (panel (e)). The
frequency distribution is a Gaussian with zero mean and unit variance.
The figure shows data obtained from numerical integration of the
dynamics and from the time-averaged theory discussed in
Section~\ref{sec:tat}.}
\label{fig:tat-comparison}
\end{figure}

We use Eq.~(\ref{eq:r-tat-final}) to obtain the behaviour of the order
parameter $r$ versus $\sigma$ for various values of $K$ and compare with that obtained from
direct numerical integration of the dynamics~(\ref{eq:eom2}) for a
lattice of size $L=3200$, see Fig.~\ref{fig:tat-comparison}. The values of $K$ are: $K=0.1$ (panel (a)), $K=0.04$ (panel (b)), $K=-0.04$
(panel (c)), $K=-0.1$
(panel (d)), and $K=-0.21$ (panel (e)). The data
have been averaged over several frequency realizations. Note that the
time-averaged theory described above is valid in the synchronized phase.
For positive as well as low negative $K$, the order parameter behaviour obtained from the theory
is in very good agreement with numerics, see
Fig.~\ref{fig:tat-comparison}, panels (a), (b) and (c). With $K$ becoming
more negative so that one approaches the tricritical point (see
Fig.~\ref{fig:parameter_space}), the deviation
between theory and numerics becomes evident, especially close to the
phase transition point, see Fig.~\ref{fig:tat-comparison}(d). For $K$
values for which one has a first-order transition, the match between the
theory and numerical results worsens substantially, even somewhat deep
into the synchronized phase, see Fig.~\ref{fig:tat-comparison}(e). 
Nevertheless, the remarkable agreement in the case
of continuous transitions lets us conclude that there is good enough merit
in using the time-averaged theory in obtaining the behavioral trend of
stationary $r$ in the synchronized phase. We anticipate that in parameter
regimes of first-order transitions, the local field set up by the
nearest-neighbor interaction competing with the global mean-field leads
to enhanced fluctuations neglected in our time-averaged theory. It would
be interesting to formulate a theory that would explain the variation of
$r$ for $K$ values for which $r$ shows a first-order transition as well as for
$K$ values to the right of the tricritical point as the latter is
approached from the side of continuous transition, see
Fig.~\ref{fig:parameter_space}. One crucial issue would then be to
devise a suitable measure that is analytically
tractable and yet able to take into account local fluctuations. A
possibility is that the one-oscillator distribution function that was employed in the
time-averaged theory is dispensed with, and one considers instead, e.g., a
two-oscillator distribution function that gives the joint probability
density for two consecutive-site oscillators
to observe given angle values at a given time instant.

\begin{figure}[]
\hspace{-0.52 cm}
\includegraphics[scale=0.22]{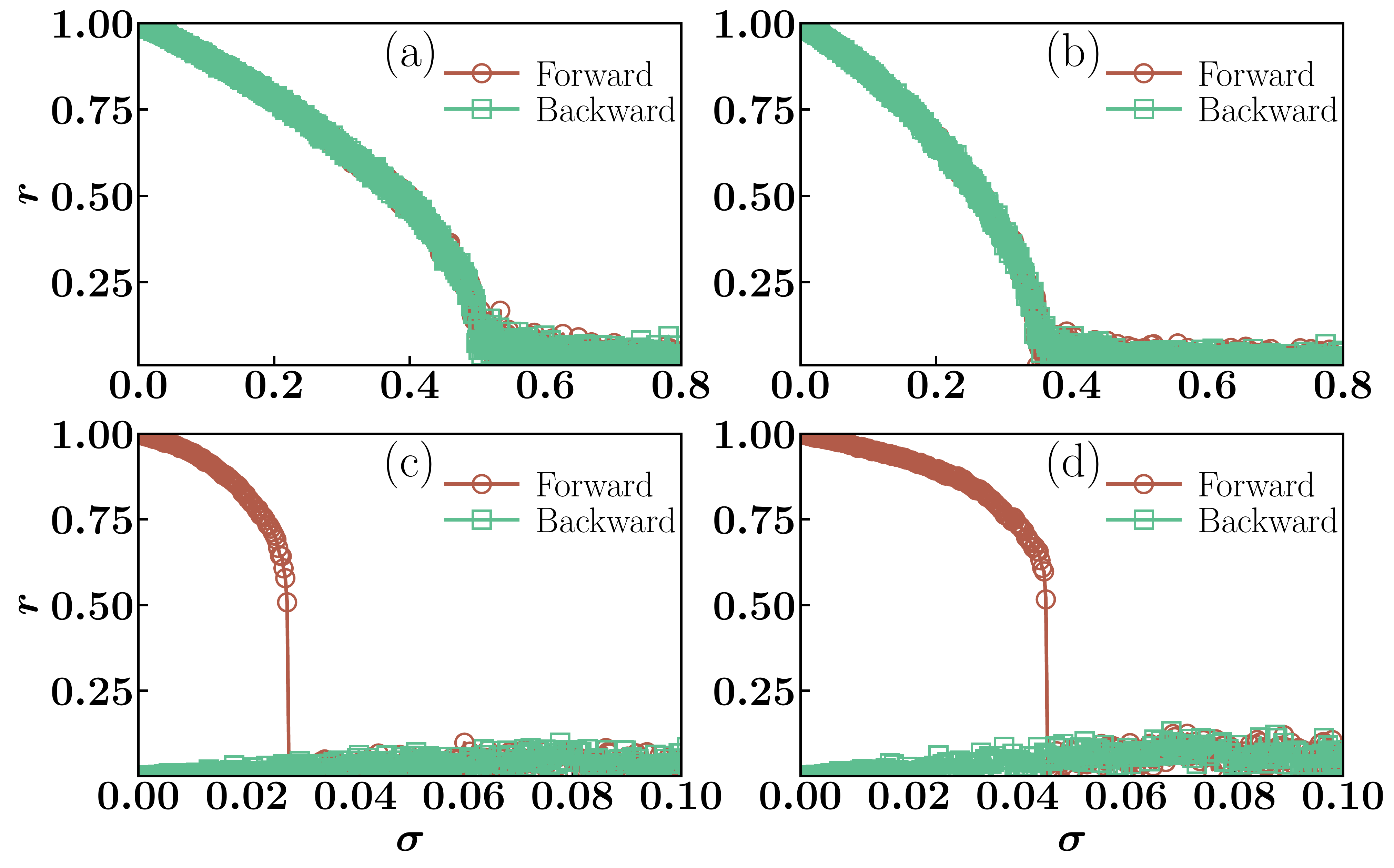}
\caption{(Color online) Variation of order parameter $r$ with
adiabatically-tuned $\sigma$ in the stationary state of the
dynamics~(\ref{eq:eom2}) on a
lattice of size $L=3200$ and
for four values of $K$, namely, $K=0.04$ (panel (a)), $K=-0.1$ (panel
(b)), $K=-0.23$ (panel (c)), and $K=-0.24$ (panel (d)). The frequency distribution $g(\omega)$ is a
Lorentzian with zero mean and unit width. The results correspond to a
typical realization of the frequencies. Hysteresis behaviour is observed only in panels
(c) and (d). The data are obtained from numerical
integration of the dynamics~(\ref{eq:eom2}).} 
\label{fig:hysteresis-lor}
\end{figure}

\section{Conclusions}
\label{sec:conclusions}

In this work, we studied a variation of the celebrated Kuramoto
model of spontaneous collective synchronization, by including in the
dynamics a nearest-neighbor interaction on a one-dimensional lattice
with periodic boundary conditions. For unimodal and symmetric frequency distributions,
we demonstrated that the resulting
dynamics exhibits a rich phase diagram in the stationary state, with the
system exhibiting synchronized and incoherent phases separated by
transition lines that could be either continuous or first-order. The
first-order and continuous transition lines meet at a tricritical point.
For such frequency distributions, the usual Kuramoto model that has only
mean-field interaction exhibits  continuous transitions and the model with solely nearest-neighbor
interactions exhibits the incoherent phase with no transitions. Our
work highlights that a competition between the two types of interactions brings in new
features, namely, that the system in contrast to the
only-nearest-neighbor case does
exhibit global synchrony, and moreover, that transitions between the
synchronized and the incoherent phase can be either continuous or
first-order depending on parameter regimes. Although we have studied in
detail the case of Gaussian frequency distributions, we have verified
for another choice of the distribution, namely, a Lorentzian, the
existence of continuous and first-order transitions, see
Fig.~\ref{fig:hysteresis-lor}. In the light of the results presented
here in the context of the model~(\ref{eq:eom1}) that is a special case
of the dynamics~(\ref{eq:HMF-eom-can-omega}) discussed in
Appendix~\ref{app0}, it would be interesting to
study the phase diagram of the latter model that is more general. Investigations in these directions will be reported elsewhere.

\appendix

\section{Motivating the form of the dynamics~(\ref{eq:eom1})}
\label{app0}

The dynamics~(\ref{eq:eom1}) may be motivated from a completely
different perspective than that of coupled oscillators, which serves to
rationalize the physical setting of the model. To this end, consider a
system of interacting rotors occupying the sites of a one-dimensional
periodic lattice of $L$ sites, with sites labeled $i=1,2,\ldots,L$. Let
$(\theta_i, p_i)$ be the canonically-conjugate variables for the $i$-th
rotor; here, the angle $\theta_i \in [0,2\pi)$, with $\theta_{i+L}=\theta_i$, is the generalized
coordinate, while $p_i$ is the corresponding conjugate momentum. The
Hamiltonian of the system is given
by~\cite{Campa2006,Dauxois2010,Campa2014}
\begin{eqnarray}
&&H=\sum_{i=1}^L \frac{p_i^2}{2I}+\frac{J}{2L}\sum_{i,j=1}^L
\left[1-\cos(\theta_i-\theta_j)\right] \nonumber\\
&&~~~~~~~~~~~~~~-K\sum_{i=1}^L
[\cos(\theta_{i+1}-\theta_i)+\cos(\theta_{i-1}-\theta_i)],
\label{eq:HMF-H}
\end{eqnarray}
which models two kinds of interactions between the rotors: a
nearest-neighbor interaction with coupling $K$ that can be either
positive or negative, and a mean-field
ferromagnetic interaction with coupling $J>0$. Here, $I$ is the common
moment of inertia of the rotors. The model~(\ref{eq:HMF-H}) naturally 
arises in the context of a class of layered magnets (such as
$\mathrm{(CH_3NH_3)_2CuCl_4}$) that in specific temperature ranges and for certain sample
shapes is faithfully described by a microscopic Hamiltonian reducible to a Hamiltonian of classical rotators on a
one-dimensional lattice with both a nearest-neighbor and a mean-field
interaction, namely, of the form~(\ref{eq:HMF-H}), see
Refs.~\cite{Sato2004,Wruvel2005,Sato2005,Dauxois2010}. Such a reduction is
supposed to be generic for systems dominated by dipolar
forces~\cite{Bramwell2009,Landau1960}, and so the Hamiltonian~(\ref{eq:HMF-H}) is not just a model of
academic interest but is strongly grounded in the physics of layered
magnetic structures.

The dynamics of the system~(\ref{eq:HMF-H}) is generated by the Hamilton's equations of
motion derived from the Hamiltonian~(\ref{eq:HMF-H}), as 
\begin{eqnarray}
&&\frac{{\rm d}\theta_i}{{\rm d}t}=\frac{p_i}{I}, \nonumber \\
\label{eq:HMF-eom}
&&\frac{{\rm d}p_i}{{\rm d}t}=\frac{J}{L}\sum_{j=1}^L
\sin(\theta_j-\theta_i)+K\sum_{j \in nn_{i}}\sin(\theta_{j}-
\theta_{i}).
\end{eqnarray}
With $K=0$, the model~(\ref{eq:HMF-H}) reduces to a paradigmatic model
of long-range interactions, the so-called Hamiltonian mean-field (HMF)
model, which has been extensively studied over the years to exemplify a
number of peculiar static and dynamic properties exhibited by long-range
interacting systems~\cite{Campa2014}. The dynamics~(\ref{eq:HMF-eom}) conserves total
energy of the system and as such models time evolution within a
microcanonical ensemble. In order to mimic the interaction of the system
with the external environment modeled as a heat bath at a constant temperature
$T$, one introduces in the spirit of Langevin dynamics a suitable
friction term in the dynamics resulting in the following time evolution within a
canonical ensemble:
\begin{eqnarray}
&&\frac{{\rm d}\theta_i}{{\rm d}t}=\frac{p_i}{I}, \nonumber \\
\label{eq:HMF-eom-can} \\
&&\frac{{\rm d}p_i}{{\rm d}t}=-\gamma \frac{p_i}{I} +\frac{J}{L}\sum_{j=1}^L
\sin(\theta_j-\theta_i)+K\sum_{j \in nn_{i}}\sin(\theta_{j}-
\theta_{i})+\eta_i(t). \nonumber 
\end{eqnarray}
Here, $\gamma>0$ is the friction constant, while $\eta_i(t)$ is a Gaussian, white noise with properties
\begin{equation}
\langle \eta_i(t)\rangle=0,~~\langle \eta_i(t)\eta_j(t')\rangle=2\gamma
k_BT \delta_{ij}\delta(t-t'),
\end{equation}
where angular brackets denote averaging with respect to noise
realizations, and $k_B$ is the Boltzmann constant. 

From the form of the Hamiltonian~(\ref{eq:HMF-H}), it is clear that the
interaction terms may induce (depending on the relative magnitudes of $J$
and $K$) a clustering of rotor angles and consequently
a macroscopic order in the system. It is then natural to define the
so-called (complex) magnetization order parameter 
\begin{equation}
me^{{\rm i}\psi}\equiv
\frac{1}{L}\sum_{j=1}^L e^{{\rm i}\theta_j},
\label{eq:HMF-m}
\end{equation}
with $m$ denoting the
magnetization or the amount of clustering present in the system at any
time instant.
Both the
dynamics~(\ref{eq:HMF-eom}) and~(\ref{eq:HMF-eom-can}) allow a
stationary state that is an equilibrium one, namely, microcanonical
equilibrium for the former and canonical equilibrium for the latter.
The
phase diagram of the model in both microcanonical and canonical
equilibrium has been studied in the past, and it has been found that
the model with $K=0$ exhibits a continuous phase transition between a
magnetized ($m \ne 0$) and a non-magnetized ($m=0$) phase at the critical
temperature $k_BT_c = J/2$ in canonical equilibrium and at the
corresponding critical energy $\epsilon_c= 3J/4$ in microcanonical
equilibrium. With $K \ne 0$, the model exhibits a very rich phase
diagram with both first-order and continuous phase transitions and a
tricritical point~\cite{Campa2006,Dauxois2010}.

Being rotors, it is natural that they may be subject to external torques
that vary from one rotor to the other. To model this situation, we may
modify the dynamics~(\ref{eq:HMF-eom-can}) to read
\begin{eqnarray}
&&\frac{{\rm d}\theta_i}{{\rm d}t}=\frac{p_i}{I}, \nonumber \\
\label{eq:HMF-eom-can-omega} \\
&&\frac{{\rm d}p_i}{{\rm d}t}=\omega_i-\gamma \frac{p_i}{I}+\frac{J}{L}\sum_{j=1}^L
\sin(\theta_j-\theta_i)\nonumber \\
&&~~~~~~~~~~~~+K\sum_{j \in nn_{i}}\sin(\theta_{j}-
\theta_{i})+\eta_i(t), \nonumber 
\end{eqnarray}
where $\omega_i$ is a quenched random variable denoting the external
torque acting on the $i$-th rotor. We may consider all the $\omega_i$'s
to be sampled from a common distribution. The
dynamics~(\ref{eq:HMF-eom-can-omega}) does not derive from an underlying
Hamiltonian, since the presence of $\omega_i$'s does not allow an
interparticle potential to be defined that is periodic in the
$\theta_i$'s (see the discussion preceding
Eq.~(\ref{eq:eom-overdamped})), and this is but natural as the $\omega_i$'s represent
after all torques applied externally to the system. An immediate
consequence is that the dynamics~(\ref{eq:HMF-eom-can-omega}) has a
stationary state that is generically a nonequilibrium one, in contrast to
the case with $\omega_i=0~\forall~i$ when as argued above the stationary
state is an equilibrium one. As opposed to equilibrium stationary states that are time-reversal invariant, encoded in the so-called
principle of detailed balance that such states satisfy, nonequilibrium
stationary states (NESSs) manifestly violate detailed balance leading to
nonzero loops of probability current in the configuration space, and  offer an active area of research in the arena
of modern day statistical mechanics~\cite{Livi2017}. Unlike equilibrium states that may all be characterized in terms
of the well-founded Gibbs-Boltzmann ensemble theory encompassing
microcanonical and canonical ensembles, a general tractable framework built in the same vein
that allows to study NESSs on a common footing is as yet lacking, implying that NESSs need to be studied
on a case-by-case basis. It is then evidently of interest to study model systems with NESSs which are
simple enough to allow for detailed analytical characterization and yet are general enough to capture the
essential features of observed physical phenomena.

Now, we may imagine
a situation in which the friction constant has such a high value that
the ration $I/\gamma \to 0$, and the
dynamics~(\ref{eq:HMF-eom-can-omega}) needs to be considered in the
overdamped limit. The resulting dynamics in this limit is obtained
from Eq.~(\ref{eq:HMF-eom-can-omega}) as
\begin{equation}
\gamma\frac{{\rm d}\theta_i}{{\rm d}t}=\omega_i+\frac{J}{L}\sum_{j=1}^L
\sin(\theta_j-\theta_i)+K\sum_{j \in nn_{i}}\sin(\theta_{j}-
\theta_{i})+\eta_i(t).
\end{equation}
Dividing throughout by $\gamma$, and redefining the couplings as
$J/\gamma \to J$, $K/\gamma \to K$ and the torque as $\omega_i \to
\omega_i/\gamma$, one obtains
\begin{equation}
\frac{{\rm d}\theta_i}{{\rm d}t}=\omega_i+\frac{J}{L}\sum_{j=1}^L
\sin(\theta_j-\theta_i)+K\sum_{j \in nn_{i}}\sin(\theta_{j}-
\theta_{i})+\zeta_i(t),
\label{eq:HMF-eom-can-omega-overdamped}
\end{equation}
where $\zeta_i(t)\equiv \eta_i(t)/\gamma$ satisfies $\langle
\zeta_i(t)\rangle=0,~\langle
\zeta_i(t)\zeta_j(t')\rangle=(2k_BT/\gamma)\delta_{ij}\delta(t-t')$.
Noting that the magnetization order parameter~(\ref{eq:HMF-m}) is
exactly identical to the Kuramoto synchronization order
parameter~(\ref{eq:order_para_definition}), the
dynamics~(\ref{eq:HMF-eom-can-omega-overdamped}) may be rewritten in
terms of the quantities $r$ and $\psi$, as    
\begin{equation}
\frac{{\rm d}\theta_i}{{\rm d}t}=\omega_i+J r \sin(\psi-\theta_i)+K\sum_{j \in nn_{i}}\sin(\theta_{j}-
\theta_{i})+\zeta_i(t).
\label{eq:HMF-eom-can-omega-overdamped-again}
\end{equation}
It
is then evident that the dynamics~(\ref{eq:eom1}) is a special case of
the dynamics~(\ref{eq:HMF-eom-can-omega-overdamped-again}) with $T$ set to
zero. We have thus provided a concrete rationale for the
model~(\ref{eq:eom1}) from a perspective other than that of Kuramoto
oscillators.

\section{Scaling theory of continuous transitions in equilibrium}
\label{app1}
Equilibrium continuous phase transitions are associated with a
singularity in the second derivative of the free energy,
and are observed strictly in an infinite system~\cite{Fisher1967}. While the limit of an
infinite system can be achieved in theoretical analysis, experiments and
numerical analysis invariably involve systems of finite size.
Finite-size scaling theory allows to estimate the phase transition
point, i.e., the parameter value at which a singularity occurs in an
infinite system, by analyzing the data for large but finite systems. For
our discussions of the finite-size scaling theory, consider a system with two different phases characterized by a real scalar order
parameter $\Psi$, and a continuous phase transition occurring as a function of
temperature $T$ with the system existing in an ordered phase with $|\Psi|>0$
(respectively, in a disordered phase with $\Psi=0$) at temperatures
below a critical temperature $T_c$ (respectively, at and above $T_c$).
Defining $t \equiv (T-T_c)/T_c$ and considering a system with linear
dimension $L$ (so that $N$, the number of degrees of freedom, scales as
$N \sim L^d$, with $d$ being the dimension of the embedding space), let
us denote the correlation length as $\xi(L)$, the 
order parameter as $\Psi(L)$, and consider the quantity $\chi(L) \equiv
L^d \left(\langle (\Psi(L))^2\rangle - \langle \Psi(L)\rangle^2\right)$,
measuring stationary-state fluctuations of the order parameter and
related to the zero-field susceptibility. Here, $\langle \cdot \rangle$
denotes time average
in the stationary state. Then, a continuous phase transition,
observed as $L \to \infty$, is
characterized by the divergence of the correlation length $\xi(\infty)$ at
temperatures around the critical point as $\xi(\infty) \sim
|t|^{-\nu};~t \to 0$, where $\nu$ is a
critical exponent~\cite{Fisher1967}. The critical exponent $\beta$
characterizes the behavior of $\Psi(\infty)$ close to the critical
point, as $\Psi(\infty) \sim (-t)^\beta;~t\to 0^-$. The quantity $\chi(\infty)$ is on the other
hand known to diverge as $\chi(\infty) \sim |t|^{-\gamma};~t \to 0$, where $\gamma$ is
another critical exponent. 
For large but finite $L$ and at a given $|t|
\to 0$, if one has $L \gg \xi(\infty)$, no significant finite-size effects
should be observed. On the other hand, for $L \ll \xi(\infty)$, the
system size will cut-off long-distance correlations, and hence, finite-size
rounding off of critical-point singularities is expected. It is then
reasonable to expect for small $|t|$ that the ratio $\xi(\infty)/L$ (or, equivalently, the
ratio $|t|L^{1/\nu}$) controls the behavior of $\chi,~\Psi$, etc, so that one may write under
the assumptions of the finite-size scaling theory the following scaling
forms~\cite{binder-book}: 
\begin{eqnarray}
&&\Psi(L) \sim L^{-\beta/\nu} f(|t|L^{1/\nu}), \nonumber \\
\label{eq:app-scaling-form} \\
&&\chi(L) \sim L^{\gamma/\nu}g(|t| L^{1/\nu}). \nonumber 
\end{eqnarray}
The scaling functions $f(x)$ and $g(x)$, defined with $x>0$,
behave in the limit $x\to \infty$ as $f(x) \sim x^\beta$ and $g(x)\sim x^{-\gamma}$. In the limit $x \to 0$, the functions behave as
$f(x)|_{x\to 0}\to$ constant and $g(x)|_{x\to 0}\to$ constant. Such forms
ensure that as required, in the limit $L \to \infty$ at a fixed and
small $|t|$, we
have $\Psi(\infty) \sim t^\beta$ and $\chi(\infty) \sim |t|^{-\gamma}$.
On the other hand, at a fixed $L$, as $|t| \to 0$, one has $\Psi(L) \sim
L^{-\beta/\nu}$ and $\chi(L) \sim L^{\gamma/\nu}$. 

In order to estimate the critical point of a continuous transition, one
proceeds as follows. For finite $L$, the infinite-$L$ divergence in $\chi$
is rounded and shifted over a finite range of temperature around a
pseudo-critical point $T_c(L)$; in the limit $L \to \infty$, the region
shrinks to zero and $T_c(L)$ converges to infinite-$L$ value $T_c$
as~\cite{binder-proc}
\begin{equation}
T_c(L)-T_c \propto L^{-1/\lambda_T},
\label{eq:TcL}
\end{equation}
with $\lambda_T$ a phenomenological exponent to characterize the
shifting of $T_c(L)$ with $L$. In numerics, one uses the data for the
maximum of $\chi(L)$ for different $L$ to obtain $T_c(L)$ as a function
of $L$. Fitting the plot to a power law of the form~(\ref{eq:TcL}) then allows to estimate $T_c$. Using this value of $T_c$ and the scaling
forms~(\ref{eq:app-scaling-form}), one then plots the
finite-$L$ data ($L^{\beta/\nu}\Psi(L)$ vs. $|t|L^{1/\nu}$ and
$L^{-\gamma/\nu}\chi(L)$ vs. $|t|L^{1/\nu}$) and obtains estimates of the critical exponents by requiring that
the data for large $L$ collapse onto each other. 
\acknowledgments
We would like to thank HPCE, IIT Madras for providing us with computing facilities in the VIRGO Super cluster. S.G. acknowledges support from the Science
and Engineering Research Board (SERB), India under SERB-TARE scheme Grant No.
TAR/2018/000023 and SERB-MATRICS scheme Grant No. MTR/2019/000560. He also thanks ICTP -- The Abdus Salam International Centre for Theoretical Physics,
Trieste, Italy for support under its Regular Associateship scheme. 

\bibliography{References_mrinal}

\end{document}